\documentstyle[epsf,twocolumn,prb,aps]{revtex}

\begin{document}
\draft

 \twocolumn[\hsize\textwidth\columnwidth\hsize\csname @twocolumnfalse\endcsname

\title{Dynamical properties of the one-dimensional Holstein model}
\author{Chunli Zhang, Eric Jeckelmann \cite{byline}, and Steven R. White}
\address{
Department of Physics and Astronomy,
University of California,
Irvine, California 92697
}
\date{\today}
\maketitle
\begin{abstract}
The spectral weight functions and the optical conductivity of
the Holstein model are studied on a one-dimensional six-site lattice 
with periodic boundary conditions
for three different electron concentrations:
a single electron, two electrons of opposite spins, and half filling.
A density matrix approach is used to obtain an optimal phonon basis
and to truncate the phonon Hilbert space without significant loss
of accuracy.
This approach allows us to calculate spectral functions for electrons
dressed locally by the optimal phonons as well as for bare electrons.
We obtain evidence for a smooth crossover 
from quasi-free electrons to an heavy itinerant small polaron 
(single-electron case) or bipolaron (two-electron case)
as the electron-phonon coupling strength increases.
At half filling we observe a crossover from a quasi-free-electron
ground state to a quasi-degenerate Peierls charge-density-wave
ground state for a finite electron-phonon coupling.
This crossover is marked by an abrupt drop of the Drude weight 
which is vanishingly small in the Peierls phase.

\end{abstract}

 ]

\section{Introduction}

The Holstein model\cite{model} has been used for many years to study
physical problems related to the electron-phonon interaction,
such as the formation of polarons and bipolarons by self-trapping
of charge carriers, or the existence of charge-density-wave (CDW)
ground state due to the Peierls instability.~\cite{pei}
While our knowledge of the ground state of this model has considerably
progressed for the past few years, our understanding of its dynamical
properties is still very limited and often disputed.
The lack of reliable results is especially important in the
non-adiabatic and intermediate electron-phonon coupling regimes,
where most of the interesting physics occurs,
such as self-trapping crossover and metal-insulator transition.
Currently, there is no well-controlled analytical method to study
these regimes and most reliable results come from numerical simulations,
such as exact 
diagonalizations~\cite{ran,mel,wel,feh,wei,ale,mar93,mar95,cap,rob97,zha,wan},
quantum Monte Carlo (QMC) simulations~\cite{rae,kor,hir,ross},
and recent density matrix renormalization group (DMRG) 
calculations~\cite{jec98,bur,jecnew}.
Among these various numerical methods, only 
exact diagonalizations can easily be used to compute 
dynamical properties of the system.
Although this technique can only be applied to small clusters due to
restrictions on computer resources, it often allows us to gain a
useful insight into the physics of the system.
Moreover, if the error due to the necessary truncation of 
the phonon Hilbert space is negligible,
this method provides numerically exact results, 
which can be used to assess the accuracy 
of other analytical or numerical methods.

In this paper we report our study of the dynamical properties of a
six-site one-dimensional Holstein lattice with periodic boundary
conditions. 
We consider three different electron concentrations: a single electron,
two electrons of opposite spins and a half-filled band (six electrons
with zero total spin).
The Holstein model describes non-interacting electrons coupled
to dispersionless phonons.
Its Hamiltonian is 
\begin{eqnarray}
H =  \Omega \, \sum_i b^\dag_i b_i 
- \gamma \, \sum_i \left (b^\dag_i + b_i \right ) 
n_i
\nonumber \\ - t \sum_{i\sigma} \left (c^\dag_{i+1\sigma}
c_{i\sigma} + c^\dag_{i\sigma} c_{i+1\sigma} \right )  \, ,
\label{eq:ham}
\end{eqnarray}
where $c^\dag_{i\sigma}$($c_{i\sigma}$) creates (annihilates) an
electron with spin $\sigma$ on site $i$,
$n_i = c^\dag_{i\uparrow} c_{i\uparrow}+c^\dag_{i\downarrow}
c_{i\downarrow}$, and 
$b^\dag_i$ and $b_i$ are
creation and annihilation operators of the local phonon mode.
The model parameters are the hopping integral $t$, 
the electron-phonon coupling constant $\gamma$, and
the bare phonon frequency $\Omega$.
For all the results presented in this paper,
the phonon frequency is chosen to be in the non-adiabatic regime 
$\Omega = t $ and we study the variations of the system properties when the 
electron-phonon coupling $\gamma$ goes from zero to the strong-coupling
regime $\gamma>\Omega, t$.

We perform exact diagonalizations of the Holstein Hamiltonian
using the efficient local phonon Hilbert space reduction method
that we have recently introduced.~\cite{zha}
This approach uses the information contained in a reduced density matrix
to generate an optimal phonon basis which allows us to truncate
the phonon Hilbert space without significant loss of accuracy.
In our previous work, this method has been demonstrated on the ground state 
of the six-site Holstein model at half filling.
Here, we extend this approach to different band fillings and 
to excited state calculations.
We also show how to use the optimal phonon basis 
to dress electrons with local phonons.
Using the Lanczos algorithm~\cite{dag},
single-particle and pair spectral functions are calculated
for bare electrons and dressed electrons,
and the Drude weight and optical conductivity
are computed.

This paper is organized as follows: in the next section, we describe our
method to obtain an optimized phonon basis and to dress electrons
with these phonons, and introduce the dynamical quantities
we have calculated. In Sec. III we present our results for the six-site
Holstein model. Finally, Sec. IV contains our conclusions.

\section{Methods}

\subsection{Optimal phonon basis}

In order to perform an exact diagonalization of the 
Hamiltonian~(\ref{eq:ham}),
one needs to introduce a finite basis to describe the phonon degrees
of freedom. 
If one uses a bare phonon basis 
(the basis made from the lowest eigenstates of the operators $b^\dag_i b_i$),
the number of phonon levels needed 
for an accurate treatment
can be quite large in the strong-coupling regime.
However, this number can be strongly reduced
by using an optimal basis (a basis that minimizes the error due to 
the truncation of the phonon Hilbert space).
In a previous work~\cite{zha} we have introduced a density matrix 
approach for generating an optimal phonon basis.
The key idea of this approach is identical to the key idea of 
DMRG\index{DMRG}\cite{whi}:
in order to eliminate states from a part of a system without loss
of accuracy, one should transform to the basis of 
eigenvectors of the
reduced density matrix\index{density matrix}, 
and discard states with low probability.
To be specific, consider any wave function $|\psi\rangle$
in the Hilbert space of the Holstein model.
Let $\alpha$ label the four possible electronic states 
of a particular site (empty, occupied by a single electron of spin
up or down, or occupied by two electrons of opposite spins)
and let $n$ label the bare phonon levels of this site.
Let $j$ label the combined states of all of the rest of the sites.
Then $|\psi\rangle$ can be written as
\begin{equation}
|\psi\rangle = \sum_{\alpha,n,j} \psi_{\alpha n,j}
|\alpha\rangle |n\rangle |j\rangle \; .
\label{eq:wfn}
\end{equation}
The reduced density matrix $\rho$ of the state $|\psi\rangle$
for this site is
\begin{equation}
\rho = \sum_{\alpha} \left [ |\alpha\rangle \langle\alpha| \otimes  
\left ( \sum_{n,r} \rho^\alpha_{n,r} |n\rangle \langle r| 
\right ) \right ] \; ,
\label{eq:rho}
\end{equation}
where $r$ is another index labeling the bare phonon levels.
This density matrix is always diagonal for the electronic
states because of the conservation of the number of electrons.
The phonon density matrix for each electronic state $\alpha$ of 
the site is given by 
\begin{equation}
\rho^\alpha_{n,r}
    = \sum_{j} \psi_{\alpha n,j} \psi_{\alpha r,j}^{*} \; .
\label{eq:rhoa}
\end{equation}
Let $w_{\alpha k}$ be the eigenvalues and $\phi_{\alpha k}(n)$
the eigenvectors of $\rho^\alpha_{n,m}$, 
where k labels the different eigenstates for a given
electronic state of the site. 
The states
\begin{equation}
|\phi_{\alpha k}\rangle = \sum_{n} \phi_{\alpha k}(n) |n\rangle 
\, , \, k=1,2,...
\label{eq:optim}
\end{equation}
form a new basis of the phonon Hilbert space for each electronic
state $\alpha$ of the site.
The $w_{\alpha k}$ are the probabilities of the site
being in the state
\begin{equation}
|\alpha,k\rangle = |\alpha \rangle |\phi_{\alpha k}\rangle 
\label{eq:optim2}
\end{equation}
if the system is in the state $|\psi\rangle$. 
These states $|\alpha,k\rangle$ form a new basis of the Hilbert space
associated with each site.
If $w_{\alpha k}$ is negligible, then the
corresponding state $|\alpha,k\rangle$ can be discarded from the
basis for the site, without affecting the state $|\psi\rangle$. 
If one wishes to keep a limited number of phonon states $m$ for a site, 
then the best states to keep corresponds to the $m$ eigenstates of 
$\rho$ with 
largest eigenvalues for each electronic state of the site. 
The corresponding phonon states $|\phi_{\alpha k}\rangle$
form an optimal phonon basis.
In the Holstein model we have found that
keeping $m = 3-5$ optimal states $|\phi_{\alpha k}\rangle$
per phonon mode for each electronic
state of the site gives results as accurate
as with hundred or more bare phonon states per site
for a wide range of parameters.

Unfortunately, in order to obtain the optimal phonon states, 
we need the target state~(\ref{eq:wfn}), for instance, the ground state.
Usually do not know this state -- we want the optimal
states to help get it.
This problem can be circumvented in several ways~\cite{zha}.
Here, we describe the algorithm that we have used to obtain 
optimal phonon bases for the ground state and low-lying states
of the Holstein model.
First, we calculate a large optimal phonon basis 
in a two-site Holstein system with appropriate parameters.
In such a small system we can carry out calculations
with enough bare phonon levels 
to render completely negligible
errors due to the truncation of the phonon Hilbert space.
Thus, target states can be obtained directly by
diagonalization in the bare phonon basis.
Then, the optimal phonon states of the two-site system are used as
the initial basis states for calculations on larger lattices.
The simplest of algorithms described in Ref.~13 is
used for the six-site system.
A single site (called the big site) contains a large number of 
the optimal phonon states obtained in the two-site systems 
(up to a few hundreds).
Each other site of the lattice is allowed to have a much smaller 
number of optimal phonon states, $m \sim 3 - 5$,
for each electronic state of the site. 
Initially these states are also
optimal phonon states of the two-site system.
The ground state of the Hamiltonian (\ref{eq:ham}) is calculated
in this reduced Hilbert space by exact diagonalization.
Then, the density matrix (\ref{eq:rho}) of the big site 
is diagonalized. 
The most probable $m$ eigenstates for each electronic state of 
the big site
form new optimal states which are used on all of the other sites 
for the next diagonalization.
These new phonon states are now optimized for the six-site system
and thus are different from the optimal
states of the two-site system.
After the first diagonalization, the new optimal states
of the six-site system are not very accurate.
Thus, diagonalizations of the Hamiltonian~(\ref{eq:ham}) and 
of the density matrix are repeated until 
the optimal states have converged.
In each diagonalization, the big site always has 
a large number of phonon states, so that it can
generate improved optimal states for the next iteration. 
After full convergence of the optimal phonon basis,
the error made by using 3-5 optimal states instead of
hundreds of bare levels is negligible: typically, the error
in the ground state energy
is smaller than $10^{-5}$ with 3 or more optimal states
when only the ground state is targeted.

To study dynamical properties we can extend the above approach by targeting
the ground state $|\psi_0 \rangle$ as well as some low-lying excited states 
$|\psi_s \rangle, s > 0$.
In this case, the density matrix of the big site is formed by adding the 
density matrices $\rho_s$ of each state 
$|\psi_s \rangle$,
with weighting factors $a_s$ (normalized by $\sum a_s = 1$),
\begin{equation}
\rho = \sum_{s}a_s \rho_s.
\label{eq:rhos}
\end{equation}
Thus, information from several states  can be included
to select the optimal phonon basis.
The weighting factors allow us to vary the influence of each state
$|\psi_s\rangle$ in the formation of the optimal phonon
basis.
Not surprisingly, we have found that more optimal states must be kept
on each site to reach a given accuracy when several states are targeted
than when only the ground state is targeted.
Obviously, these additional optimal states are necessary to describe
accurately the excited states $|\psi_s\rangle$.
However, they do not seem to be necessary to obtain a qualitative 
description of dynamical properties
(see the discussion in Sec.~\ref{sec:comp}). 
Therefore, we usually target only the ground state in our
calculations.

\subsection{Dressing of electronic operators}

A very interesting feature of our method is that it provides a very simple
way to dress electrons with phonons.
Let assume that the density matrix eigenstates 
$\phi_{\alpha k}(n), k=1,2,...$ are ranked
by decreasing weight $w_{\alpha k}$ 
for each electronic state $\alpha$.
For a given index $k$, the weights $w_{\alpha k}$ and eigenstates 
$\phi_{\alpha k}(n)$ of the different electronic states $\alpha$ 
often seem completely unrelated.
However, we can consider the relative weight 
\begin{equation}
r_{\alpha k} = \frac{w_{\alpha k}}{\sum_{q} w_{\alpha q}}
\end{equation}
of an optimal state $|\phi_{\alpha k}\rangle$ compared to the weight
of all states in the corresponding optimal basis.
As noted previously, $w_{\alpha k}$ (and thus $r_{\alpha k}$)
decreases rapidly with $k$. 
We have found that, for a given index $k$ the variations of
$r_{\alpha k}$ as a function of the electronic state $\alpha$ 
are much smaller (at least by one order of magnitude) 
than the variations of $r_{\alpha k}$ between successive values of $k$.
Therefore, there is an unambiguous  relation
between optimal phonon states  
$|\phi_{\alpha k}\rangle$ for different electronic occupations of
a site $\alpha$ given by their relative weights $r_{\alpha k}$. 
This one-to-one mapping between optimal states can be used to
dress electronic operators.

All electronic operators can be written as the sum and product of operators
acting on a single site, for instance $c^\dag_{i\sigma}$ and $c_{i\sigma}$. 
Such a local operator is diagonal in the bare phonon basis 
and can be written as
\begin{equation}
O = \left ( \sum_{\alpha,\beta}
O_{\alpha,\beta} \, |\alpha\rangle \langle\beta| \right )
\otimes I_{ph} \, ,
\end{equation}
where $\alpha$ and $\beta$ label the four possible electronic states 
of the site
and $I_{ph}$ is the identity operator acting in the Hilbert space
of the local phonon mode.
Now we can define the corresponding dressed operator as
\begin{equation}
\tilde{O} = \sum_{\alpha,\beta}
\Bigl( \, O_{\alpha,\beta} \, \,  |\alpha\rangle \langle\beta| 
\otimes U_{\alpha,\beta} \, \Bigr) \, ,
\label{eq:dressing}
\end{equation}
where $U_{\alpha,\beta}$ is a unitary operator in the {\it phonon}
Hilbert space given by
\begin{equation}
U_{\alpha,\beta} = \sum_{k} |\phi_{\alpha k}\rangle 
\langle\phi_{\beta k}| \; .
\label{eq:unitary}
\end{equation}
Obviously, for $\alpha=\beta$ we have $U_{\alpha,\beta}=I_{ph}$
because the density matrix eigenstates $\phi_{\alpha k}(n)$ satisfy
the orthonormalization condition
$\sum_k \phi_{\alpha k}(n) \, \phi_{\alpha k}(r) = \delta_{n,r}$.
However, for $\alpha\neq\beta$, $U_{\alpha,\beta}$ is not trivial
because the eigenstates for different electronic states
are unrelated, in general.
Nevertheless, $U_{\alpha,\beta}$ is unambiguously defined
thanks to the one-to-one  mapping between eigenstates
discussed above.
To be rigorous, the operator $U_{\alpha,\beta}$
is unitary only if we sum up the
index $k$ in Eq.~(\ref{eq:unitary})
over the infinite number of states in the 
basis~(\ref{eq:optim}).
As we generally know only $m \sim 3-5$ optimal phonon states,
the operator $U_{\alpha,\beta}$ also involves a projection onto
the subspace spanned by these few states.
However, by construction all wave functions that we calculate 
have a negligible weight out of this subspace. 
Thus, $U_{\alpha,\beta}$ can be regarded as a unitary transformation
for all practical purposes.

Clearly, the operator $\tilde{O}$ transforms electronic
states as the bare operator does, but it also transforms the
phonon degrees of freedom accordingly. 
For instance, the operator $\tilde{c}_{i,\sigma}$ not only removes
an electron with spin $\sigma$ form the site $i$, but it also
transforms the phonon mode on this site, changing optimal phonon states
for a site with two electrons into optimal states for a site
with an electron of spin $-\sigma$ and changing optimal states
for a site with one electron of spin $\sigma$ into the optimal states
of an unoccupied site.
Therefore, the operator $\tilde{O}$ acts on electrons dressed
by the local optimal phonon states as the bare operator acts on bare
electrons.
For instance, the operator $\tilde{c}^\dag_{i,\sigma}$ creates
a dressed electron of spin $\sigma$ on the site $i$.
We note, however, that the dressing of electrons by phonons at a 
finite distance 
from the electrons is completely neglected with this method.

In two cases ($\gamma=0$ and $t=0$) it is possible to
calculate the optimal phonon basis analytically and thus
to understand the transformation~(\ref{eq:dressing}).
In the weak coupling limit ($\gamma/t, \gamma/\Omega \rightarrow 0$)
the optimal phonon states resemble the bare phonon
levels for each electronic state of the site.
Therefore, the unitary transformation $U_{\alpha,\beta}$
is similar to the identity operator for any values of
$\alpha$ and $\beta$ and thus, $\tilde{O} \approx O$.
In the strong-coupling anti-adiabatic limit ($\gamma, \Omega >> t$)
the optimal phonon states are simply the eigenstates
of quantum oscillators with an equilibrium position shifted
by $2\gamma/\Omega N_\alpha$,  
where $N_\alpha$ is the number of electrons on the site in the 
electronic state $\alpha$, 
$\;n_i \,|\alpha\rangle = N_\alpha |\alpha\rangle$.
This corresponds to the states obtained by applying the
Lang-Firsov unitary transformation~\cite{lf}
\begin{equation}
S(g) = e^{-g \sum_i (b_i^\dag - b_i) n_i} \; 
\label{eq:lfop}
\end{equation}
with $g=\gamma/\Omega$ to the bare states.
Therefore, in this limit we have 
\begin{equation}
\tilde{O} = S(\gamma/\Omega) \; O \; S^{-1}(\gamma/\Omega)
\label{eq:lf}
\end{equation}
and the electronic operators obtained
with the transformation~(\ref{eq:dressing})
are completely equivalent to those defined
in other works~\cite{mel,rob97,rob98}
using the Lang-Firsov transformation.
However, in the general case,
the transformation~(\ref{eq:dressing}) is more accurate than
the Lang-Firsov transformation.
The later is only an (analytical) approximation
of the transformation of a local phonon mode 
as a function of the electronic occupation of a site,
while the former is based on a (numerically) exact transformation
of the phonon states.

\subsection{Dynamical quantities}

We compute dynamical properties such as spectral weight
functions and optical conductivity using the 
Lanczos algorithm combined with the continued fraction method.\cite{dag}
This algorithm yields not only the dynamical correlation and response
functions of the system, but also the most important eigenstates 
$|\psi_s\rangle$ that contribute to these functions and that we need
to build the density matrix~(\ref{eq:rhos}).
We define the spectral weight function as
\begin{equation}
A(p, \omega) = \frac{1}{\pi} 
Im \left [ \langle \psi_0| c_{p\sigma}^\dag \frac{1}
{H - \omega -E_{0}-i\epsilon} c_{p\sigma} |\psi_0\rangle \right ] \, ,
\label{eq:spectral}
\end{equation}
where $|\psi_0\rangle$ is the ground state wave function for a given number 
of electrons, $E_0$ is the ground state energy, and 
the operators $c_{p\sigma}^\dag$ and $c_{p\sigma}$ create
and annihilate an electron with momentum $p$ and spin $\sigma$, respectively.
In all the results presented in this paper, we have used a broadening
factor $\epsilon=0.1$.
The total weight of this spectral function (obtained by integrating
over $\omega$) is equal to the momentum density distribution
\begin{equation}
n_\sigma(p) = \langle \psi_0| c^\dag_{p \sigma} c_{p \sigma} |\psi_0\rangle .
\label{eq:total}
\end{equation}
We also define a spectral function $\tilde{A}(p,\omega)$
and its total weight $\tilde{n}_\sigma(p)$,
where we substitute dressed operators $\tilde{c}_{p\sigma}^\dag$
and $\tilde{c}_{p\sigma}$ for the corresponding bare operators
in (\ref{eq:spectral}) and (\ref{eq:total}).
When there are more than one electron in the lattice, one expects
electrons to form tightly bound pairs at strong electron-phonon coupling.
Thus, it is interesting to study the pair spectral function
\begin{equation}
P(p,\omega) = \frac{1}{\pi} 
Im \left [ \langle \psi_{0}| \Delta_{p}^\dag 
\frac{1}{H  - \omega - E_{0} - i\epsilon} 
\Delta_{p} |\psi_{0}\rangle \right ] ,
\label{eq:spectral2}
\end{equation}
where 
\begin{equation}
\Delta^\dag_p  =  \frac{1}{\sqrt{N}} \sum_{j} e^{ipj} 
c^\dag_{j\uparrow} c^\dag_{j\downarrow}
\end{equation}
and its hermitian conjugate $\Delta_p$ 
are pair operators with momentum $p$,
and $N$ being the number of sites.
The total weight of this spectral function is given by
\begin{equation}
d(p) = \langle \psi_0| \Delta^\dag_p \Delta_p |\psi_0\rangle .
\label{eq:total2}
\end{equation}
In this case too, we introduce
a spectral function $\tilde{P}(p,\omega)$ and its total weight
$\tilde{d}(p)$ using dressed electronic 
operators $\tilde{\Delta}^\dag_p$ and $\tilde{\Delta}_p$
in Eqs.~(\ref{eq:spectral2}) and (\ref{eq:total2}).

The real part of the optical conductivity is made up of a 
Drude peak at $\omega=0$ and an incoherent part for $\omega>0$, 
$\sigma(\omega) = D\delta (\omega) + \sigma'(\omega)$.
The incoherent part of the conductivity is given by
\begin{equation}
\sigma'(\omega)  
 =  \frac{e^2}{\omega N}
    Im \left [ \langle \psi_{0}| J^\dag \frac{1}{H 
    -E_{0}-\omega-i\epsilon} J |\psi_{0}\rangle \right] ,
\end{equation}
where the current operator is defined as
\begin{equation}
J =  {it} \sum_{j\sigma}  
(c^\dag_{j+1\sigma} c_{j\sigma} - c^\dag_{j\sigma} c_{j+1\sigma}).
\end{equation}
$\sigma'(\omega)$ can be calculated using the Lanczos method,~\cite{dag} 
then the Drude weight $D$ can be evaluated using the well-known
sum rule 
\begin{equation}
\int_0^\infty \sigma(\omega) d\omega = \frac{\pi e^2}{2} (-T)
\label{eq:totalopt}
\end{equation}
relating the total weight of the optical conductivity
to the electronic kinetic energy per site
\begin{equation}
T=\frac{-t}{N}\sum_{j\sigma}
\langle \psi_0 | c^\dag_{j+1\sigma} c_{j\sigma} 
+ c^\dag_{j\sigma} c_{j+1\sigma} | \psi_0 \rangle.
\end{equation}
In this paper, the spectral functions and the incoherent part of
the conductivity are always shown in arbitrary units.
Quantitative results for the Drude weight $D$ are expressed in units
of $2 \pi e^2 t$.

\subsection{Comparison with other approaches}
\label{sec:comp}

As a first check of our method, 
we have compared our exact diagonalization results
for the lowest eigenstates of small Holstein clusters
(up to six sites) with DMRG calculations.
We have always found a good quantitative agreement.
For instance, the eigenenergies obtained with both methods agree
within $10^{-3}t$ for at least up to the 18th lowest eigenstates
in a six-site lattice.

We have also carried out calculations of dynamical properties
in the two-site Holstein model where we can keep enough 
phonon levels to obtain numerically exact results.
In this case, we have calculated the optimal phonon basis 
in various ways, changing both the number of targeted states 
$|\psi_s \rangle$ and the weighting factors $a_s$ to
build the density matrix~(\ref{eq:rhos}).
We have found that overlaps between corresponding optimal phonon states
in the different basis are always larger than $90\%$.
The spectral functions and conductivity calculated using the different 
basis also agree qualitatively.  
Therefore, the inclusion of excited states in the density 
matrix~(\ref{eq:rhos}) does not
seem to be necessary to obtain a qualitative description of
dynamical properties. 

Finally we have also calculated the optical conductivity of the six-site
Holstein model of spinless fermions at half filling.
We have found a satisfactory agreement with the results obtained 
recently by Weisse and Fehske~\cite{wei} using 
completely different approaches to truncate the phonon
Hilbert space and to calculate the conductivity.

The optimal phonon approach used in this work
is so efficient that we can easily 
carry out calculations for dynamical quantities
that require powerful parallel computers
when a standard phonon Hilbert space truncation method 
is used.~\cite{wel,feh,wei,wan}
All results presented here have been obtained on a
workstation with a 133MHz processor and 150Mb of RAM memory.

\section{Results}
\subsection{Single electron}
The case of the Holstein model with a single electron
is known as the polaron problem.
This case has been extensively studied with 
both analytical and numerical methods.
~\cite{ran,mel,wel,feh,ale,mar93,mar95,cap,rob97,rae,kor,jec98,rob98,opol,trugman}
For weak coupling ($\gamma/\Omega<1$
and $\gamma^2/\Omega<2t$)
the ground state is a quasi-free electron
dragging a phonon cloud. 
For strong coupling ($\gamma/\Omega > 1$ and $\gamma^2/\Omega > 2t$)
the electron becomes trapped by the
lattice distortion that it generates.
The quasi-particle composed of this self-trapped electron and the 
accompanying lattice distortion is called a polaron.
The polaron is said to be small when the spatial extension of
the self-trapped state is limited to one site.
It is known that a smooth crossover occurs from the quasi-free
electron ground state to a small polaron ground state as the 
electron-phonon coupling increases.

\begin{figure}[ht]
\epsfxsize=3.375 in\centerline{\epsffile{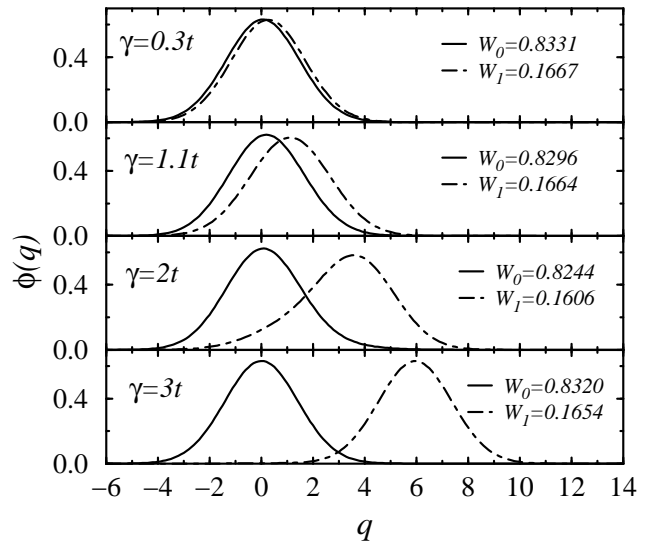}}
\caption{Single-electron system: optimal phonon wave functions $\phi$ 
as a function of the oscillator coordinate $q$ and their weights $W$ 
for the two possible occupations of a site ($N_\alpha = 0,1$)
and different electron-phonon couplings $\gamma$.
}
\label{fig:phi1}
\end{figure}

First, we examine the optimal phonon states obtained with our method.
Figure~\ref{fig:phi1} shows the optimal phonon wave functions $\phi(q)$ 
as a function of the phonon coordinate $q=b\!+\!b^\dag$ 
for different electron-phonon couplings.  
Only the most important optimal state
is shown for each of the two possible electronic occupations 
of a site ($N_\alpha=0,1$).  
For weak coupling ($\gamma=0.3t$)
the optimal states are similar to the bare phonon levels
and thus, the wave function $\phi(q)$ is just the 
ground state of a quantum oscillator.
As the coupling increases, the optimal states change smoothly
and become increasingly distinct.
For all coupling each wave function $\phi(q)$ has a large overlap
with the lowest eigenstate of a quantum oscillator with a 

\noindent
shifted
equilibrium position.
This shift is always very small for $N_\alpha=0$,  but 
for $N_\alpha=1$ it increases with increasing coupling and 
tends to $2\gamma/\Omega$ at strong coupling. 
This is in agreement with the strong-coupling theory that
predicts optimal phonon states given by the
Lang-Firsov transformation~(\ref{eq:lfop})
with $g=\gamma/\Omega$.
However, in the general case,
the optimal phonon states are different from the states obtained
with this transformation.
First, one can see in Fig.~\ref{fig:phi1} that the oscillator
shift for $N_\alpha=1$ is smaller than $2\gamma/\Omega$ for intermediate
couplings ($\gamma=1.1t$) and reaches this value only for strong coupling
($\gamma=3t$).
Of course, this difference could be taken into account simply
by using an effective parameter $g$ smaller
than $\gamma/\Omega$ in Eq.~(\ref{eq:lfop}). 
However, there are other features of the optimal states that a simple
Lang-Firsov transformation can not reproduce. 
For instance, for $\gamma = 2t$ 
the optimal wave function $\phi(q)$ for $N_\alpha = 1$
has an important tail at low $q$.
This can be understood as a retardation effect due to the finite
phonon frequency $\Omega/t$. 
Most of the time a site is unoccupied and the phonon mode is in the 
optimal state for $N_\alpha=0$. 
When the electron hops on this site, the phonon mode can not
adapt instantaneously.
Thus, its state for $N_\alpha=1$ becomes a combination
of the states obtained using the Lang-Firsov transformation
for $N_\alpha=0$ and $N_\alpha=1$.
In Fig.~\ref{fig:phi1} we also give the weight $W_0$ and $W_1$
of the most important optimal state for $N_\alpha=0$ and 
$N_\alpha=1$, respectively.
$W_1$ is much smaller than $W_0$ because the probability
of finding the electron on a given site is only 1/6 while
the probability of a site being empty is 5/6.
For $\gamma = 0$ and $\gamma \rightarrow \infty$ one can show
that there is only one optimal state for each electronic state
of a site and thus, $W_0=5/6$ and $W_1=1/6$.
For intermediate couplings, $W_0$ and $W_1$ become smaller,
showing the increasing importance of the higher optimal phonon states.

\begin{figure}[ht]
\epsfxsize=3.375 in\centerline{\epsffile{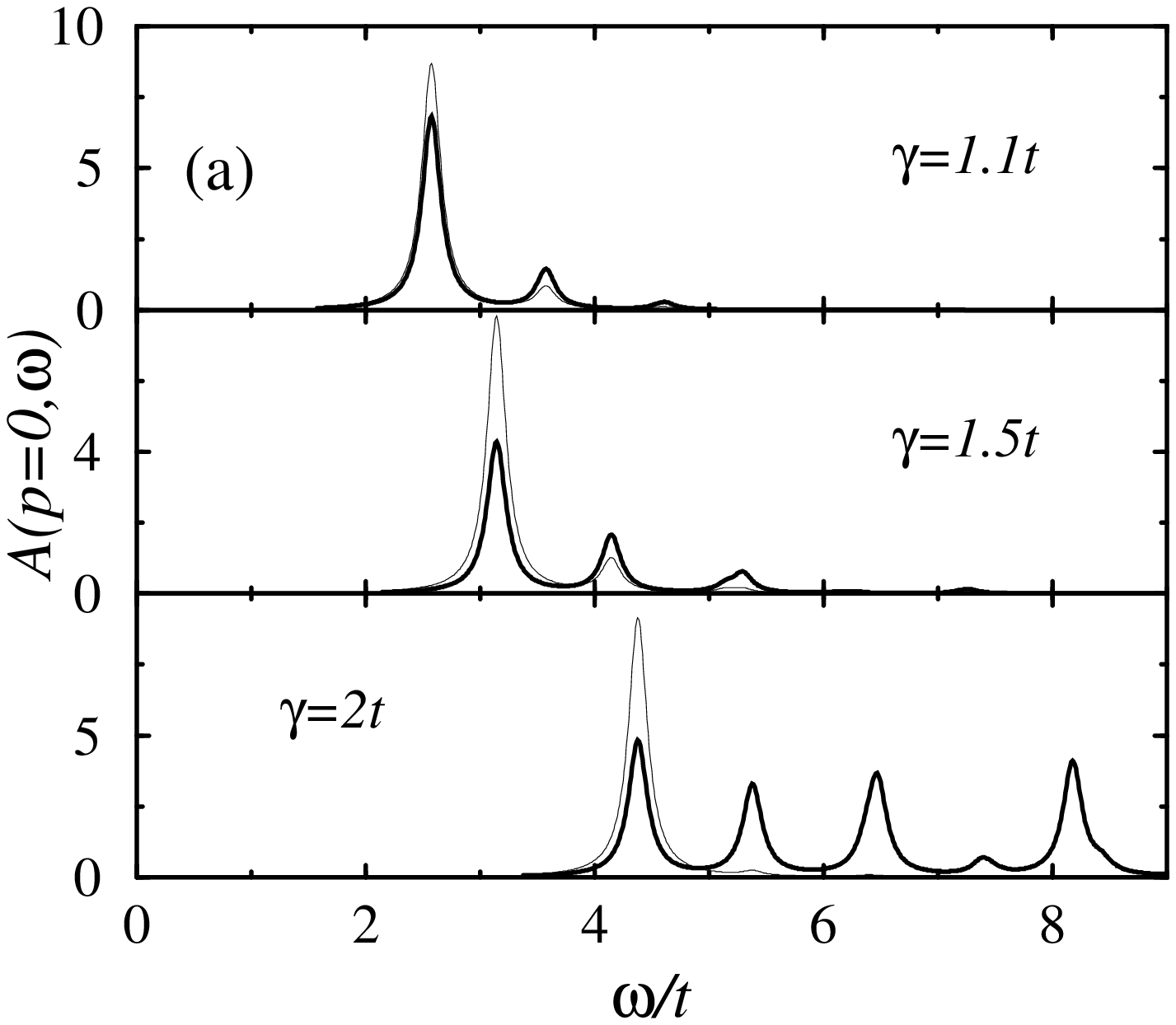}}
\end{figure}
\vspace{-7mm}
\begin{figure}[ht]
\epsfxsize=3.375 in\centerline{\epsffile{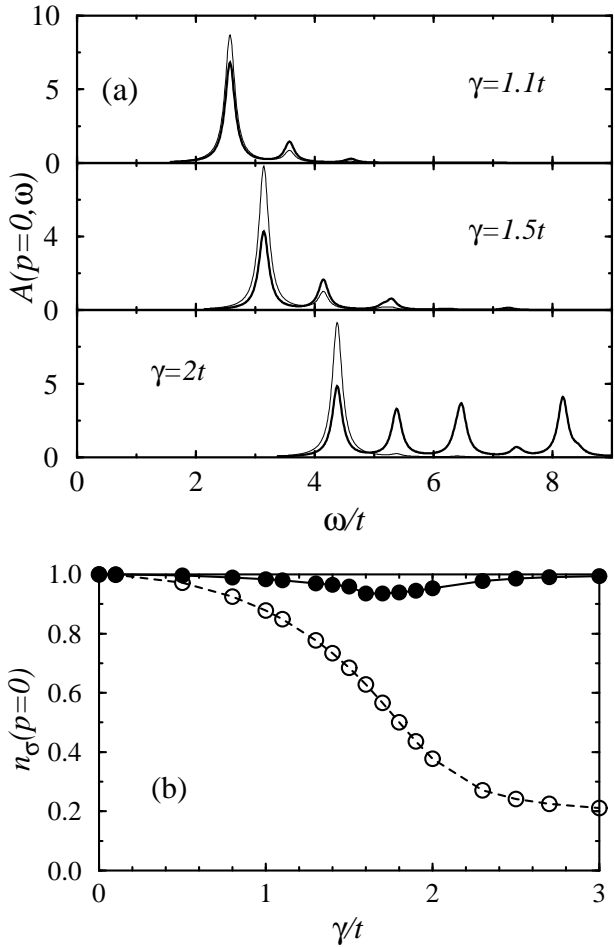}}
\caption{Single-electron system: (a) spectral functions 
$A(p\!=\!0,\omega)$ (thick line) 
and $\tilde{A}(p\!=\!0,\omega)$ (thin line)
for different electron-phonon couplings $\gamma$.
For $\gamma=2t$ $P(p\!=\!0,\omega)$
has been expanded by a factor 5.
(b) Total spectral weight $n_\sigma(p=0)$ (open circle)
and $\tilde{n}_\sigma(p=0)$ (filled circle)
as a function of the electron-phonon coupling.}
\label{fig:A1}
\end{figure}

\noindent
However, for all cases presented in Fig.~\ref{fig:phi1},
the two most important optimal states constitute more than $98\%$ of the 
total weight.

For weak electron-phonon coupling both spectral functions 
$A(p\!=\!0,\omega)$ and $\tilde{A}(p\!=\!0,\omega)$ 
have a single peak at $\omega = 2t$
because the ground state wave function is simply
\begin{equation}
|\psi_0\rangle \approx c^\dag_{p=0,\sigma} |0\rangle
\approx \tilde{c}^\dag_{p=0,\sigma} |0\rangle,
\end{equation}
where $|0\rangle$ is the vacuum state (without electron nor phonon),
with a ground state energy of about $-2t$.
Figure~\ref{fig:A1}(a) shows that
satellite peaks appear above the dominant peak energy
in both spectral functions for larger couplings.
The position of the dominant peak shift to higher energy
as the coupling $\gamma$ increases.
The distance between these peaks is roughly $\Omega$, with some peaks too 
small to be seen.  
We can easily understand the structure of these spectral functions.
Equation~(\ref{eq:spectral}) with $\epsilon \rightarrow 0$ can be written as 

\begin{equation}
A(p,\omega) = \sum_{n} | \langle \phi_n|c_{p\sigma}|\psi_0\rangle |^{2}
\delta (\omega - (\varepsilon_n -E_0)),
\end{equation}
where $E_0$ and $|\psi_0\rangle$ are the ground state energy and wave 
function for a single electron, and $\varepsilon_n$ and 
$|\phi_n\rangle$ are a complete set of energies and eigenstates for 
a lattice containing no electron and thus, only non-interacting local phonons.
In this case, all eigenenergies are of the type $\varepsilon_n = m \Omega$, 
where $m$ is an positive integer number.
Therefore, $A(p,\omega)$ and $\tilde{A}(p,\omega)$ contain only peaks 
with spacing $\Omega$ starting from $-E_0$.
In $A(p=0,\omega)$ the weight of the dominant peak shifts increasingly
to the satellite peaks and in the strong coupling regime ($\gamma=2t$), 
no dominant peak can be identified [Fig.~\ref{fig:A1}(a)].
Moreover, the total weight $n_\sigma(p=0)$ of this spectral function
decreases continuously as $\gamma$ increases [see Fig.~\ref{fig:A1}(b)]
and in the strong-coupling limit it tends to the value $1/6$ signaling
a completely localized electron.
(More precisely, if the electron is localized on a single site of the
six-site lattice then
$n_\sigma(p)=1/N=1/6$ for all momentum $p$.)
This decrease and the considerable incoherent contribution to 
$A(p=0,\omega)$ shows that the free electron state 
$c^\dag_{p=0,\sigma} |0\rangle$ is no longer a good starting point
for a description of the ground state for $\gamma \geq 1.5t$.
On the other hand, $\tilde{A}(p=0,\omega)$ contains a well defined 
quasi-particle peak for all couplings [Fig.~\ref{fig:A1}(a)], 
indicating that the motion of 
the electron is closely accompanied by a local phonon cloud represented by 
the optimal phonon states.
The position of this peak is determined by the ground state energy 
$E_0$ as explained above.
For $\gamma=2t$ 
this position gives a polaron energy that approaches the 
strong-coupling result $E_0 = -\gamma^2/\Omega$.
The total weight $\tilde{n}_\sigma(p=0)$ of this spectral function 
is very interesting [Fig.~\ref{fig:A1}(b)].
First, $\tilde{n}_\sigma(p=0)$ is always larger than 0.9, showing
that the ground state is very well described by 
\begin{equation}
|\psi_0\rangle \approx \tilde{c}^\dag_{p=0,\sigma} |0\rangle
\end{equation}
for all couplings.
Furthermore, we note that $\tilde{n}_\sigma(p=0)$ first decreases slightly as 
$\gamma$ increases and then tends to 1 in the strong-coupling limit.
We think that the initial decrease of $\tilde{n}_\sigma(p=0)$ shows the 
increasing importance of the extended phonon cloud following
the electron 
as one goes from the non-interacting limit to the intermediate
electron-phonon coupling regime.
The operator $\tilde{c}^\dag_{p=0,\sigma}$
dressed only by local phonons does not describe this extended phonon cloud.
For larger coupling $\gamma$ the phonon cloud collapses to a single site
as a small polaron is formed and thus, 
the dressed operator $\tilde{c}^\dag_{p=0,\sigma}$ becomes again 
an almost exact description of the ground state.
In this regime, 
the operators $\tilde{c}^\dag_{p, \sigma}$ and 
$\tilde{c}^\dag_{p, \sigma}$ obtained
with the transformation~(\ref{eq:dressing}) are similar
to the ``small polaron operators'' 
obtained using the Lang-Firsov transformation~\cite{mel,rob97,rob98}
as discussed in the previous section.

\begin{figure}[ht]
\epsfxsize=3.375 in\centerline{\epsffile{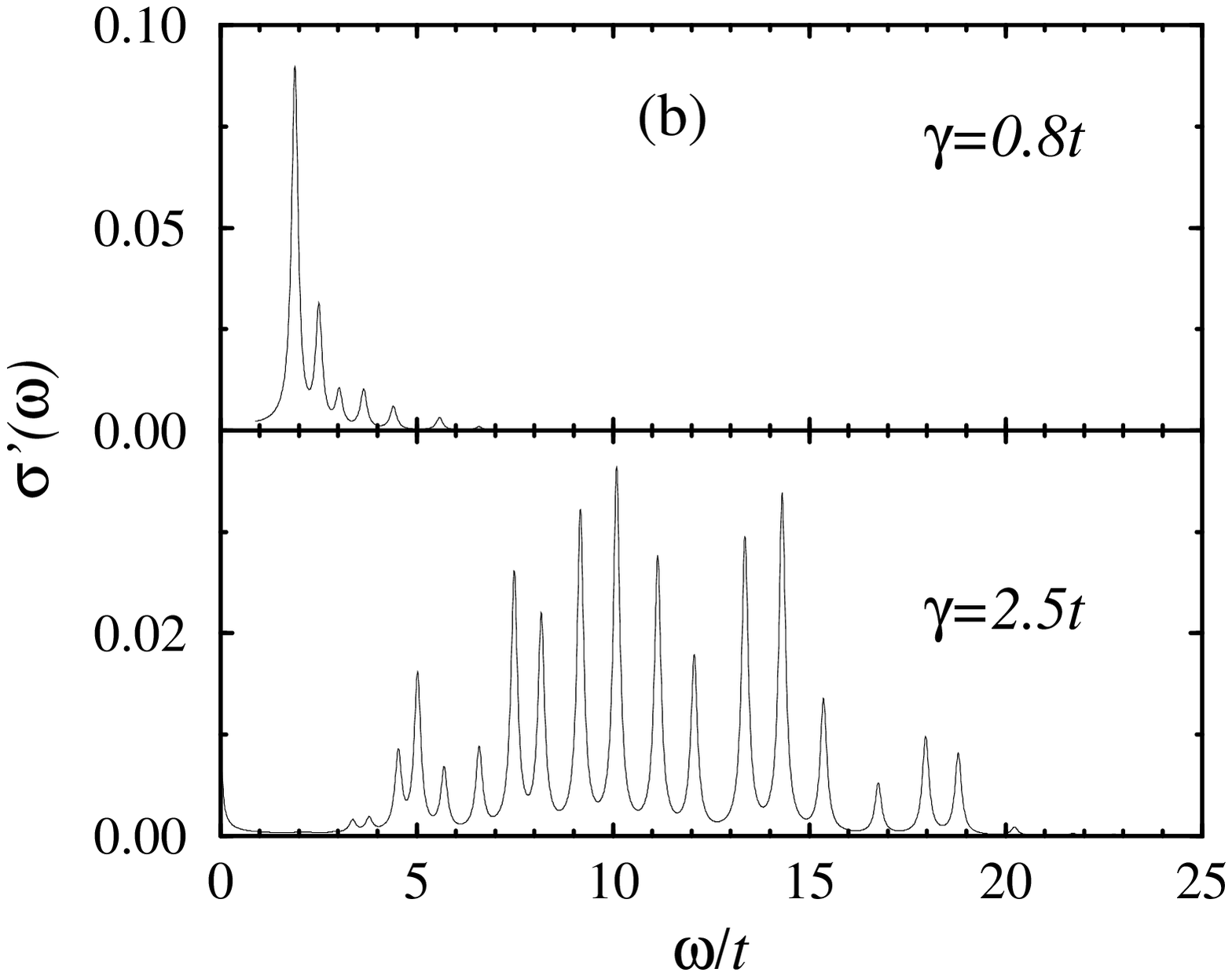}}
\caption{Single-electron system: (a) the Drude weight 
$D$ and the kinetic energy per site $T$ 
as a function of the electron-phonon coupling $\gamma$.
(b) The incoherent part of the optical conductivity $\sigma'(\omega)$
in the quasi-free electron regime ($\gamma=0.8t$) and in the small 
polaron regime ($\gamma=2.5t$).}
\label{fig:S1}
\end{figure}

The evolution of the Drude $D$ weight and of the kinetic energy
per site
$T$ is shown as a function of $\gamma$ in  Fig.~\ref{fig:S1}(a).
The kinetic energy gives the total weight of the optical conductivity
according to Eq.~(\ref{eq:totalopt}) 
while the Drude wei-

\begin{figure}[ht]
\epsfxsize=3.375 in\centerline{\epsffile{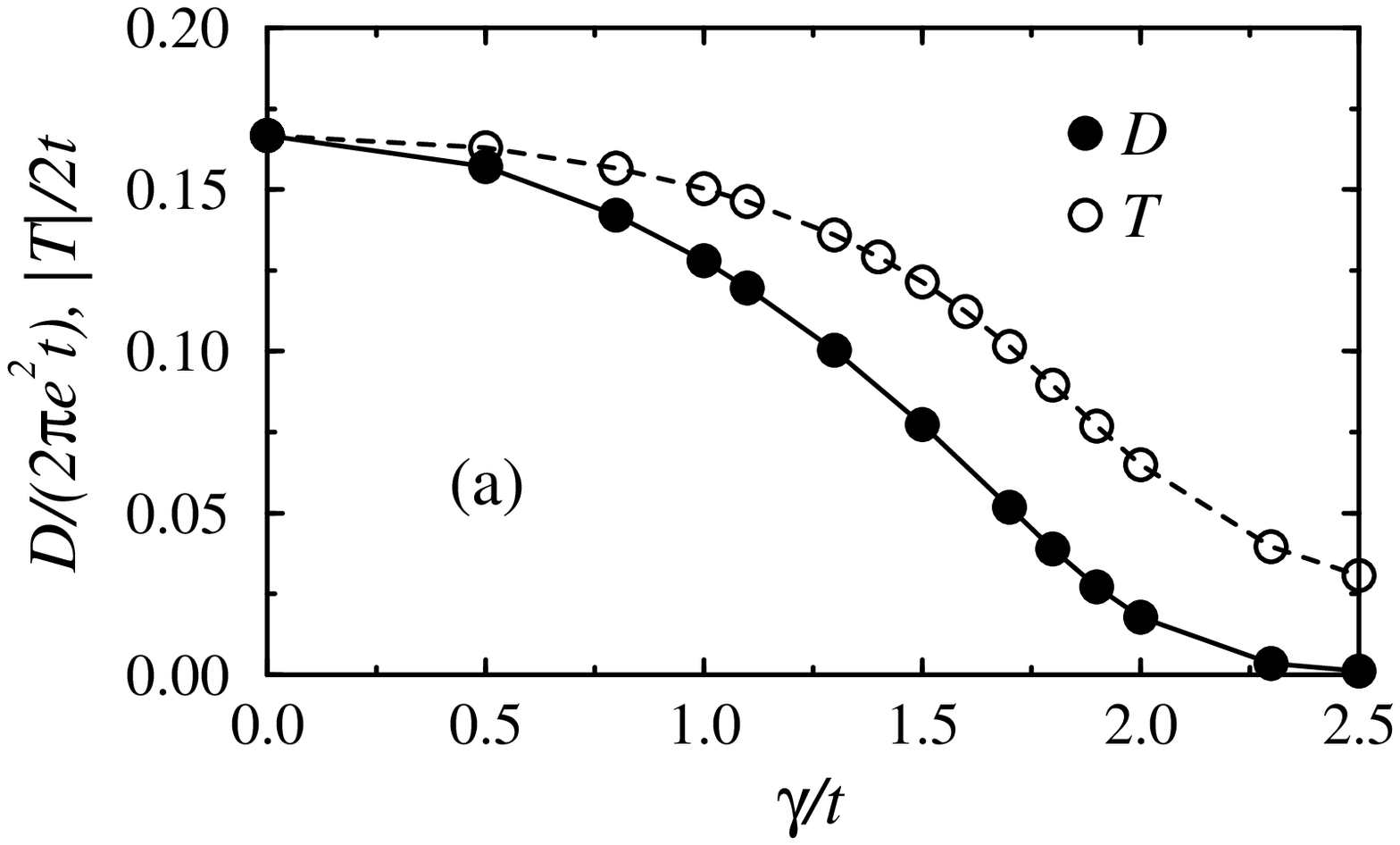}}
\end{figure}
\vspace{-8mm}

\noindent
ght measures the contribution of coherent motion
to the optical conductivity.
In Fig.~\ref{fig:S1}(a)
the units have been chosen so that both quantities appear equal
when there is no contribution from the incoherent part of the conductivity
$\sigma'(\omega)$, as in a non-interacting system ($\gamma=0$).
We see that both the Drude weight and the average kinetic energy 
decreases smoothly as the coupling increases.
However, the Drude weight decreases much faster and becomes
very small for $\gamma > 2t$.  
The slow decrease of $D$ for small $\gamma$ shows the slight increase
of the electron effective mass as it has to drag an increasingly
important phonon cloud.
The small but finite value of $D$ for large $\gamma$ reflects
the fact that a polaron moves coherently, as a free carrier,
in the Holstein model, but its effective mass is much larger than
the bare mass of the electron.
The decrease of the ratio $D/T$ implies that incoherent processes
become more important in the optical conductivity as $\gamma$
increases. 
As seen in Fig.~\ref{fig:S1}(b), $\sigma'(\omega)$ 
becomes also fairly complex in the polaronic regime ($\gamma=2.5t$).
The structure of $\sigma'(\omega)$ is different from
the optical conductivity expected for a small polaron
in the adiabatic regime~\cite{emin}.
In particular, no dominant peak is visible at the energy
corresponding to the depth of the lattice potential
which traps the electron ($\omega \approx 2\gamma^2/\Omega$).
This discrepancy is probably the consequence of the non-adiabatic
phonon frequency $\Omega = t$ used in our calculations.
In this regime, many-phonon optical excitations become as important as 
the purely electronic transition at $\omega \approx 2\gamma^2/\Omega$.
For weak electron-phonon coupling ($\gamma=0.8t$) the conductivity
$\sigma'(\omega)$ has very little weight and its structure
is mostly determined by the discrete electronic energy levels of the 
non-interacting system.~\cite{rob98}

Our results agree perfectly with the known features
of the single-electron Holstein model 
discussed at the beginning of this section.
In particular, they confirm that the ground state is formed
by an itinerant quasi-particle (quasi-free electron or polaron) for all 
electron-phonon couplings, but that the crossover 
to the polaronic regime is accompanied by a substantial enhancement
of the quasi-particle effective mass.
In the polaronic regime our results also show that the weight of 
incoherent processes is much more important than the Drude weight
in the optical conductivity.
Therefore, small perturbations which are neglected in the Holstein model,
such as disorder or thermal phonons, are likely to suppress the
coherent motion of the small polaron 
in more realistic models or actual materials.

\subsection{Two electrons with opposite spins}
The case of two electrons of opposite spins is not as well understood as
the single electron case.~\cite{ran,mel,wel,mar95}
For strong enough coupling 
electrons are trapped by the lattice distortion that they generate and form
an itinerant bound pair called a bipolaron.
If both electrons are localized on a single site,
the bipolaron is said to be small.
It is known that a small bipolaron is formed in the strong coupling limit
of the Holstein model with two electrons.
What happens at weaker coupling before the onset of small bipolaron
formation is still debated.
Three different scenarios are possible: both electrons remain free,
two independent polarons are formed, or a ``large" bipolaron is formed
(a bipolaron whose spatial extension is larger than one site).

\begin{figure}[ht]
\epsfxsize=3.375 in\centerline{\epsffile{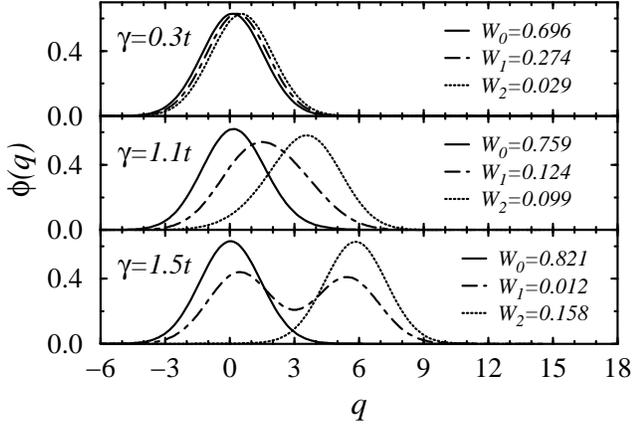}}
\caption{Two-electron system: most important optimal phonon modes 
as a function of the oscillator coordinate $q$ and their weights $W$ 
for the three possible occupations of a site ($N_\alpha = 0,1,2$)
and different electron-phonon couplings $\gamma$.
For a singly occupied site ($N_\alpha = 1$) the optimal states
are identical for an electron with spin up or down. 
The weight $W_1$ is the sum of the probability of the most important 
state for each of the two spin orientations. 
}
\label{fig:phi2}
\end{figure}

Figure~\ref{fig:phi2} shows the most important optimal phonon wave functions 
for each possible electronic occupation of a site ($N_\alpha =0,1,2$) 
for different electron-phonon couplings.
(In this system, optimal phonon states are similar if the site is 
occupied by one electron with spin up or down, so we do not distinguish 
between these two cases).
As in the single-electron system, the optimal phonon states are just
the ground state of the operator $b^\dag b$ in the weak coupling limit.
The weights of these optimal states are determined by the probability
of a given site being empty ($W_0=25/36$), occupied by one electron
($W_1=10/36 $), or by both electrons ($W_2=1/36$)
when there are two independent electrons on a six-site lattice.

\noindent
As $\gamma$ increases, $W_1$ decreases and tends to zero in the strong
coupling limit while both $W_0$ and $W_2$ increase and tend to
5/6 and 1/6, respectively. 
These weights correspond to the occupation probability of 
a particular site when the two electrons form an itinerant tightly
bound pair, as in a small bipolaron.
For intermediate couplings, the combined weight of the most probable 
phonon states in Fig.~\ref{fig:phi2} 
is smaller than 1, showing the importance
of higher  optimal phonon states, but always
remains larger than 98\% of the total weight.
The wave functions for $N_\alpha=0,2$ have large overlaps with 
the ground state of a shifted harmonic oscillator for any coupling.
For $\gamma \geq 1.5t$ this shift is very close to 
the strong-coupling theory prediction $2 N_\alpha \gamma/\Omega$.
Therefore, these optimal states can be obtained with the Lang-Firsov
transformation~(\ref{eq:lfop}) using an appropriate parameter $g$,
but $g$ reaches the theoretical value $\gamma/\Omega$ only at 
strong coupling, as in the single-electron case.
The optimal wave function for $N_\alpha=1$, however, changes
significantly as the coupling increases.
In Fig.~\ref{fig:phi2} one can see that this wave function is approximately
the superposition of the wave functions for $N_\alpha=0$ and $N_\alpha=2$
when $\gamma=1.5t$.
This wave function cannot be obtained by applying the Lang-Firsov 
transformation to a bare phonon state.
The shape of this optimal state can be understood as a retardation effect.
Most of the time a site is empty or occupied by both electrons and
the phonon modes are in the corresponding optimal states.
The electronic states with $N_\alpha=1$ are essential intermediate 
states for allowing the coherent motion of the bipolaron, as the 
Holstein Hamiltonian~(\ref{eq:ham}) contains only one-electron hopping terms,
but they have very low probability.
Electrons do not spend enough time in these states
for the phonon modes to adapt. 
Therefore, optimal phonon states for $N_\alpha=1$ are determined by the 

\begin{figure}[ht]
\epsfxsize=3.375 in\centerline{\epsffile{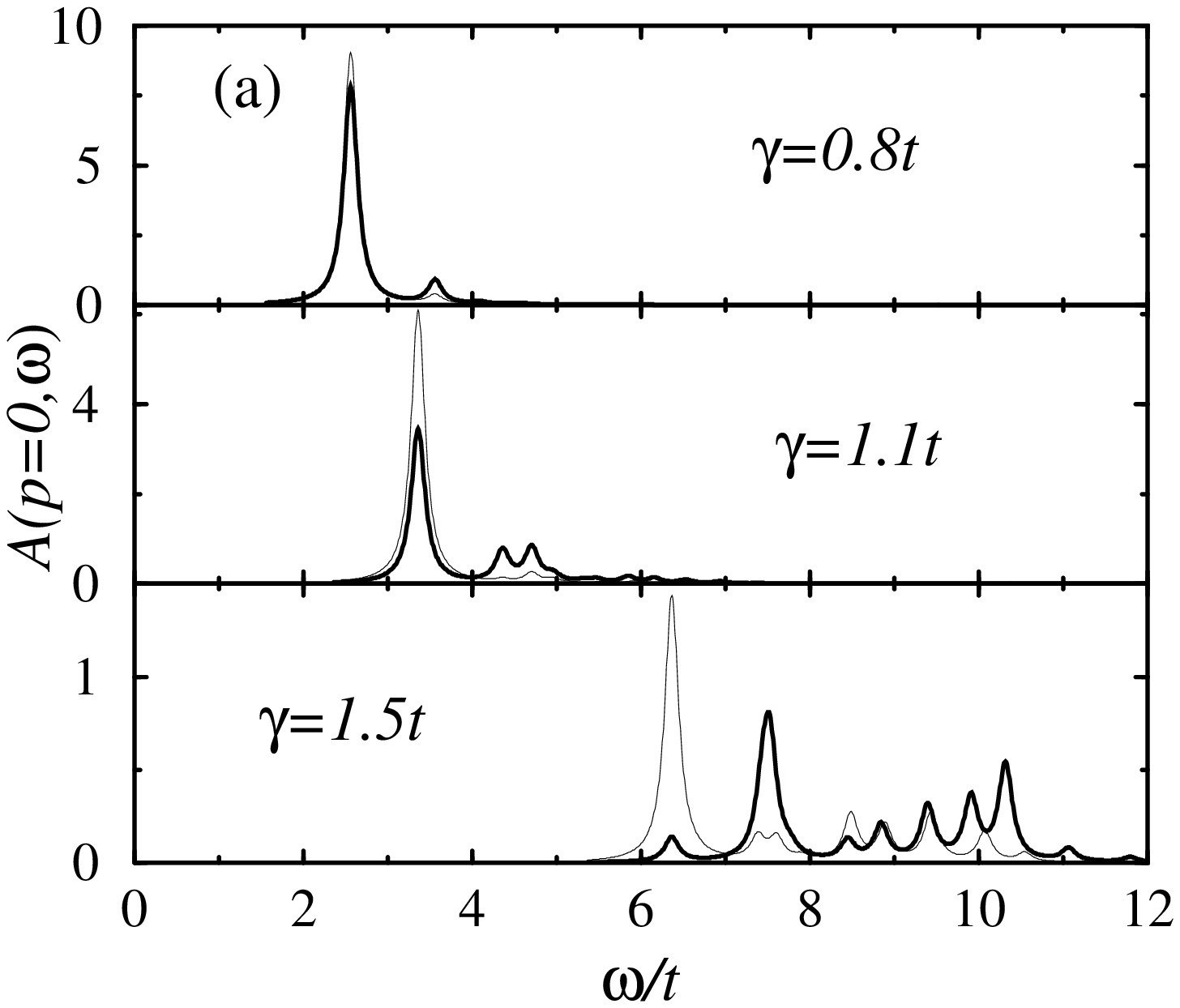}}
\end{figure}
\vspace{-7mm}
\begin{figure}[ht]
\epsfxsize=3.375 in\centerline{\epsffile{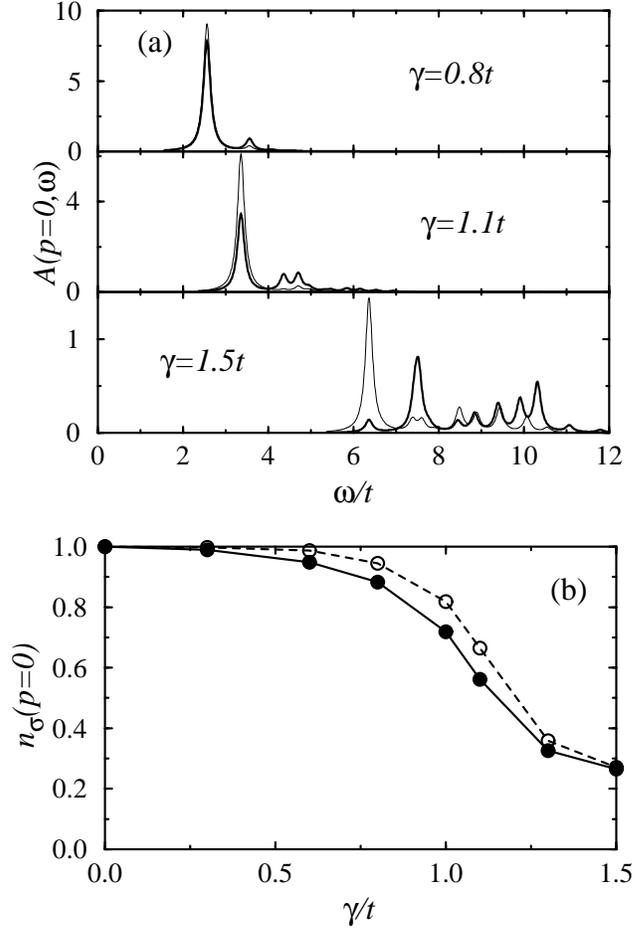}}
\caption{Two-electron system: (a) the spectral functions 
$A(p\!=\!0,\omega)$ (thick line) and $\tilde{A}(p\!=\!0,\omega)$
(thin line)
for different electron-phonon couplings $\gamma$.
(b) Total spectral weight $n_\sigma(p=0)$ (filled circle)
and $\tilde{n}_\sigma(p=0)$ (open circle)
as a function of the electron-phonon coupling.}
\label{fig:A2}
\end{figure}

\noindent
optimal phonon states for $N_\alpha=0$ and $N_\alpha=2$.

The single-particle
spectral functions $A(p\!=\!0,\omega)$ and $\tilde{A}(p\!=\!0,\omega)$
are similar to those of the single-electron system at weak-coupling.
A single peak is observed at $\omega \approx 2t$ because both electrons
occupy the lowest one-particle  eigenstates with energy $-2t$, and
the ground state wave function is  
\begin{equation}
|\psi_0\rangle \approx c^\dag_{p=0,\uparrow} 
c^\dag_{p=0,\downarrow} |0\rangle
\approx \tilde{c}^\dag_{p=0,\uparrow} 
\tilde{c}^\dag_{p=0,\downarrow} |0\rangle \; .
\end{equation}
In Fig.~\ref{fig:A2}(a) we show both spectral functions
for different electron-phonon couplings.
The results for bare electrons are qualitatively similar to those
observed in the single-electron system.
As the coupling strength increases, 
the weight of the dominate peak shifts to an increasing
number of satellites peaks until
no well-defined quasi-particle peak can be observed. 
Dressing the fermion operator simplifies the spectral weight structure a bit,
but important incoherent contributions are still observable.
Moreover, the total spectral weight becomes small 
at strong coupling for both bare and dressed electrons
[Fig.~\ref{fig:A2}(b)].

\begin{figure}[ht]
\epsfxsize=3.375 in\centerline{\epsffile{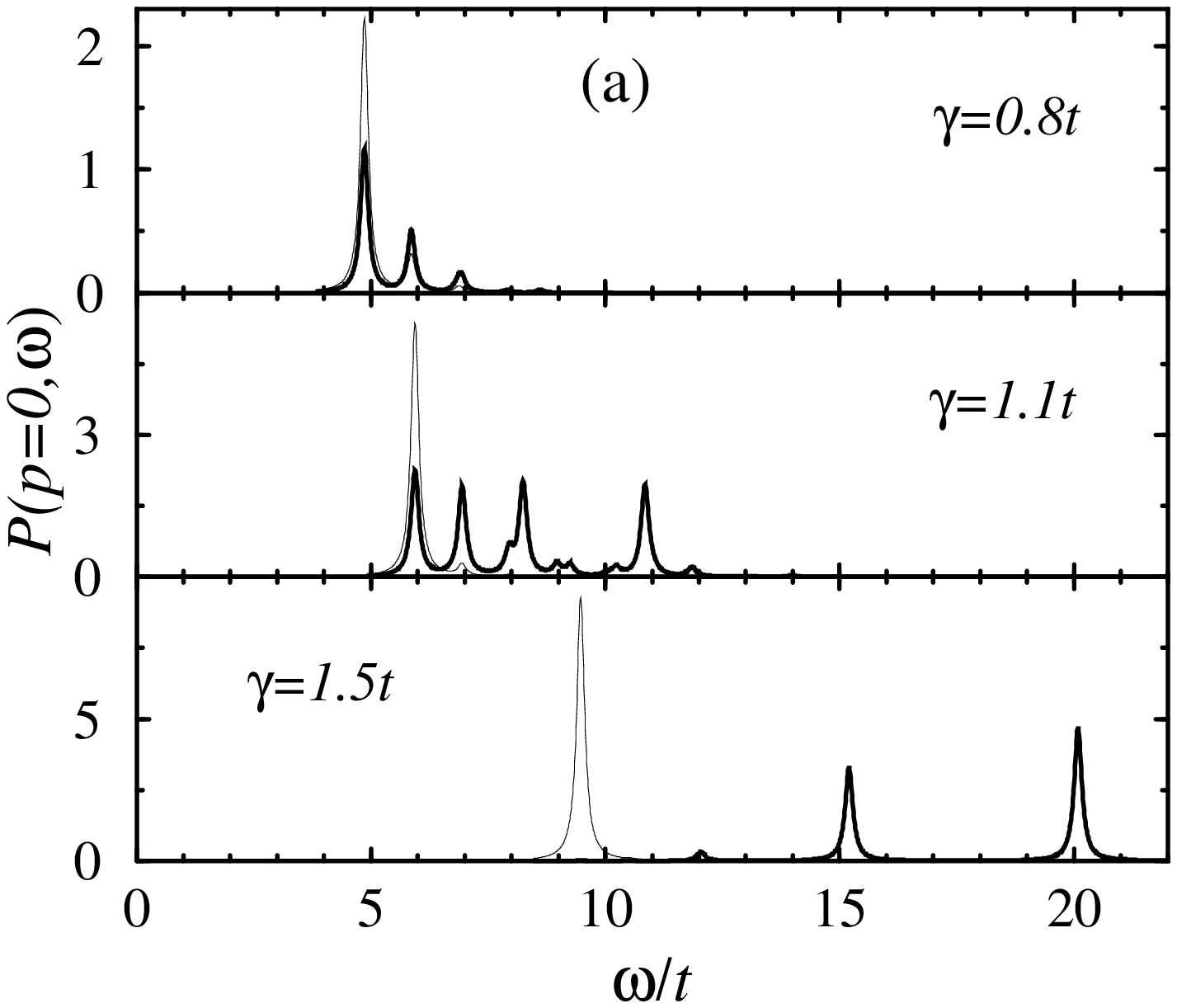}}
\end{figure}
\vspace{-8mm}
\begin{figure}[ht]
\epsfxsize=3.375 in\centerline{\epsffile{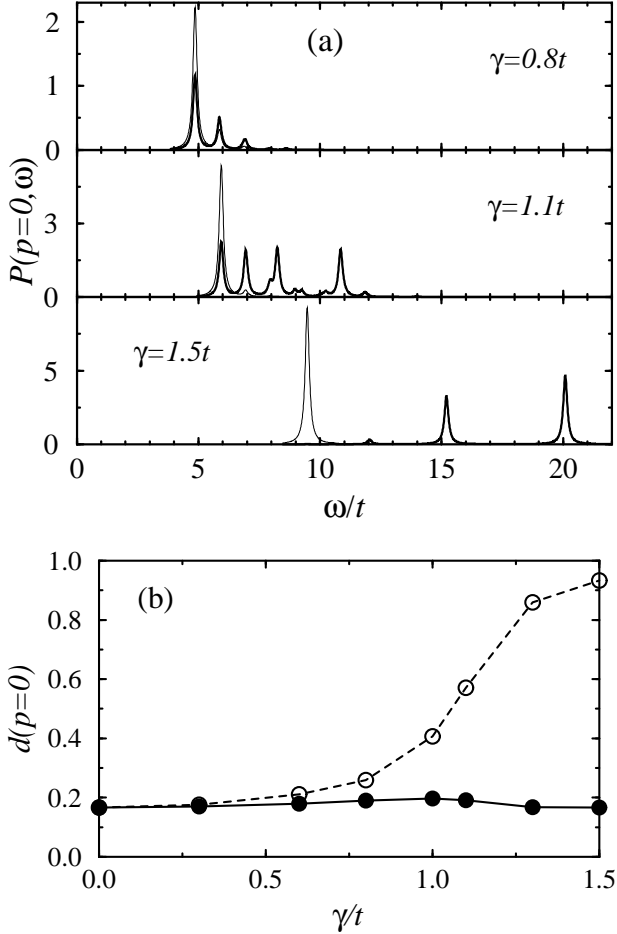}}
\caption{Two-electron system:
(a) pair spectral functions $P(p\!=\!0,\omega)$ (thick line) 
and $\tilde{P}(p\!=\!0,\omega)$ (thin line) 
for different electron-phonon couplings $\gamma$.
For $\gamma=1.1t$ and $1.5t$, $P(p\!=\!0,\omega)$
has been expanded by a factor 5.
(b) Total pair spectral weight $d(p=0)$ (filled circle)
and $\tilde{d}(p=0)$ (open circle) as a function of 
the electron-phonon coupling.}
\label{fig:P2}
\end{figure}

\noindent
Therefore, the ground state can no longer be described by two
independent electrons or quasi-particles as was done above for weak coupling.

At weak electron-phonon coupling the pair spectral 
functions $P(p=0,\omega)$ and $\tilde{P}(p=0,\omega)$
have a single peak at $\omega \approx 4t$ on a finite system,
but the weight of this peak vanishes as $1/N$
in the thermodynamic limit. 
Figure~\ref{fig:P2}(a) shows both functions
for stronger couplings. 
The evolution of the structure of $P(p=0,\omega)$ for increasing coupling
is qualitatively similar to that of the single-particle
spectral function.
The weight of the dominant peak is progressively shifted to 
satellite peaks until no dominant peak can be observed. 
However, the total spectral weight $d(p=0)$ remains almost constant 
(close to 1/N=1/6) for all couplings [Fig.~\ref{fig:P2}(b)].
As this value is a finite size effect for
$\gamma=0$, we think that the pair spectral function 
vanishes uniformly in an infinite Holstein lattice. 
The dressed pair spectral function is much more interesting
[see Fig.~\ref{fig:P2}(a)]. 
It shows a single dominant peak for all 
electron-phonon couplings,
though small satellite peaks are present at intermediate coupling.
Moreover, the total spectral weight $\tilde{d}(p=0)$ increases
with $\gamma$ and tends to 1 for strong coupling [Fig.~\ref{fig:P2}(b)].
This means that the ground state is given by 
\begin{equation}
|\psi_0\rangle \approx \tilde{\Delta}^\dag_{p=0} |0\rangle 
\end{equation}
in the strong-coupling regime.
This state describes an itinerant small bipolaron with momentum $p=0$. 
As the optimal phonon states for $N_\alpha = 0,2$ are close
to the states generated with the Lang-Firsov transformation~(\ref{eq:lfop})
for strong-enough coupling,
the dressed pair operator $\tilde{\Delta}_{p}$ is similar
to the ``bipolaron operator'' built with this transformation
in Ref.~4.
The structure of both $P(p=0,\omega)$ and $\tilde{P}(p=0,\omega)$
can be understood using the same arguments as for 
$A(p=0,\omega)$ and $\tilde{A}(p=0,\omega)$ in the single-electron case.
Therefore, $P(p=0,\omega)$ and $\tilde{P}(p=0,\omega)$ contain only
peaks with spacing $\Omega$ (with some peaks to small
to be seen) starting from $-E_0$, where $E_0$ is now
the ground state energy of the two-electron system.
For $\gamma=1.5t$, the position of the dominant peak of 
$\tilde{P}(p=0,\omega)$ shown in Fig.~~\ref{fig:P2}(a)
gives a bipolaron energy close to the strong-coupling result
$E_0=-4\gamma^2/\Omega$.

\begin{figure}[ht]
\epsfxsize=3.375 in\centerline{\epsffile{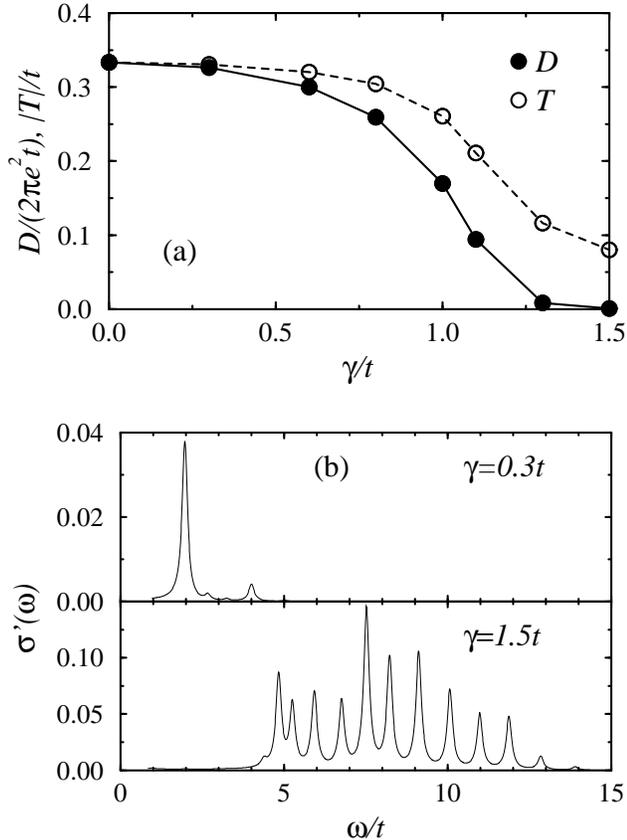}}
\caption{Two-electron system: (a) the Drude weight 
$D$ and the kinetic energy per site $T$ as a function of 
the electron-phonon coupling $\gamma$.
(b) The incoherent part of the optical conductivity $\sigma'(\omega)$
in the quasi-free electron regime ($\gamma=0.3t$) and
in the bipolaronic regime ($\gamma=1.5t$).}
\label{fig:S2}
\end{figure}

Figure~\ref{fig:S2}(a) shows 
the evolution of the Drude weight and of the kinetic energy 
per site
as a function of the electron-phonon coupling.
In this case too, the units are chosen so that both quantities are equal
in the absence of incoherent contributions to the optical conductivity.
Both quantities decrease smoothly as the coupling increases. 
As in the single-electron case, the small reduction of $D$ at weak
coupling shows the slightly renormalized effective mass of the
quasi-free electrons while the small but finite value of $D$ at 
strong coupling shows that the bipolaron is a heavy itinerant 
quasi-particle.
The diminishing ratio $D/T$ means that incoherent processes
become more important as $\gamma$ increases.
The incoherent part of the optical conductivity 
$\sigma'(\omega)$ is shown
in Fig.~\ref{fig:S2}(b) for both the quasi-free-electron regime
and the bipolaronic regime.
In the quasi-free-electron regime ($\gamma=0.3t$), 
$\sigma'(\omega)$ has a very low weight but a simple
structure which is determined by the discrete electronic levels 
of a non-interacting six-site lattice.
In the bipolaronic regime ($\gamma=1.5t$), $\sigma'(\omega)$ 
is fairly complex and the features
predicted for a bipolaron in the adiabatic regime are not 
visible.~\cite{emin}
For instance, there is no clear peak at the energy
$\omega\approx4\gamma^2/\Omega$ corresponding to the depth of
the lattice potential trapping the electrons.
As already suggested in the single-electron case, this 
is probably due to the non-adiabatic phonon frequency ($\Omega =t$) 
used in this work.

Our results show that the ground state of the two-electron system
is composed of two independent quasi-free electrons at least up to
$\gamma = 0.8t$ but it is a small bipolaron at least from $\gamma=1.3t$. 
There is a smooth crossover from one regime to the other as 
the electron-phonon coupling increases.
The nature of the ground state for $0.8t < \gamma < 1.3t$ is not 
directly determined by

\begin{figure}[ht]
\epsfxsize=3.375 in\centerline{\epsffile{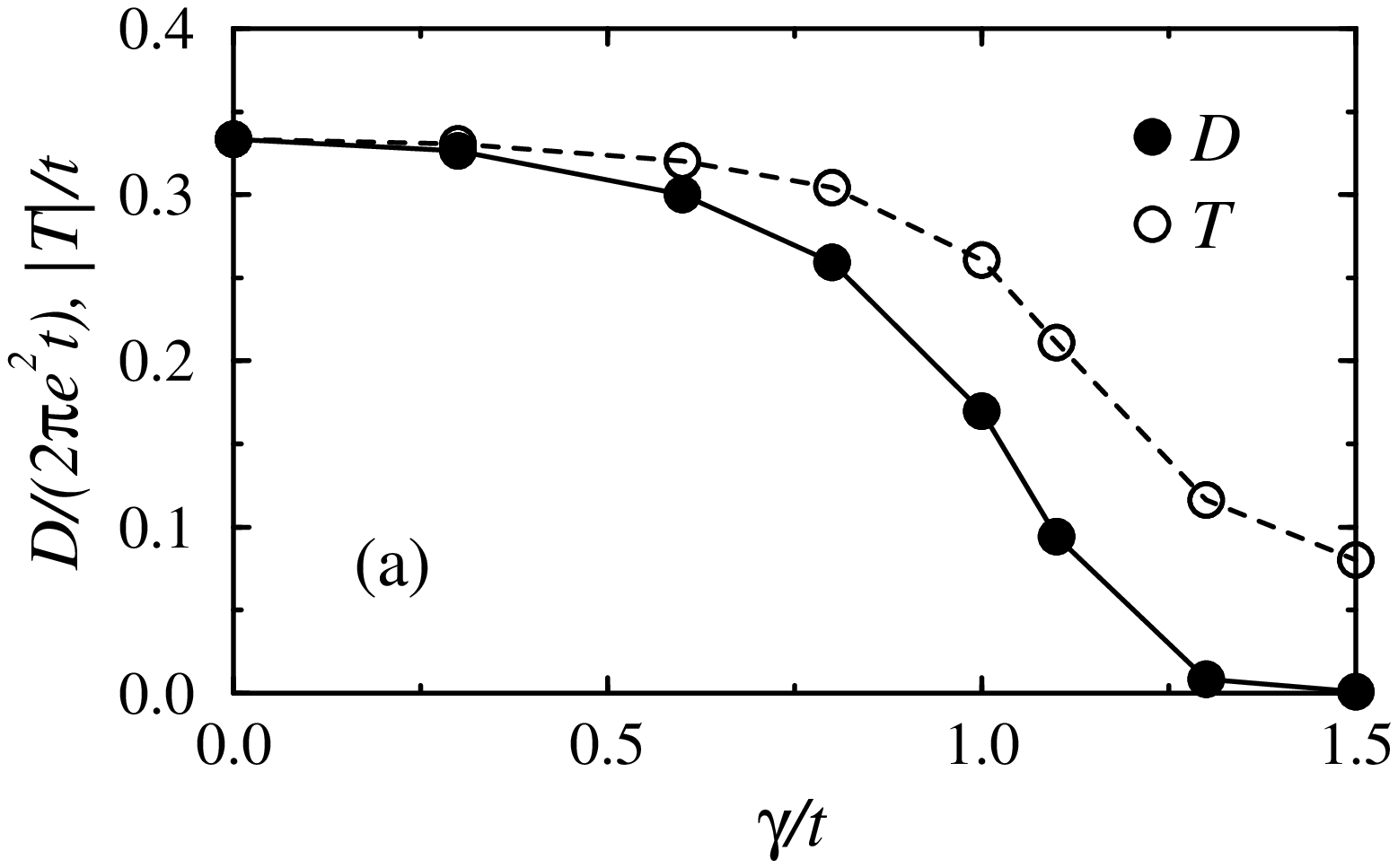}}
\end{figure}
\vspace{-7mm}

\noindent
our study.
Clearly, there is no sign of a polaron in any of the quantities we 
have calculated
and the small polaron appears at stronger coupling ($\gamma \approx 1.5t$) 
than the small bipolaron.
Thus, we can conclude that a pair of small polarons would be instable with 
respect to the formation of a small bipolaron.
The only reasonable candidate for the ground state in the intermediate
regime is the large bipolaron, but this state is difficult to study
in the small lattice considered here.
DMRG calculations for long Holstein chains confirm that the ground
state evolves smoothly from a pair of quasi-free electrons to a large 
bipolaron and then to a small bipolaron as the electron-phonon coupling 
increases.~\cite{jecun} 
As in the polaron case, the large effective mass of the bipolaron
and the dominance of incoherent processes in its optical conductivity
mean that coherent motion of a small bipolaron is unlikely in more
realistic models.

\subsection{Half filling}

Mean-field theory predicts that the ground state
of the half-filled Holstein model is a doubly degenerate 
CDW state with a dimerized lattice and a gap
at the Fermi surface for any finite electron-phonon coupling
due to the well-known Peierls instability.~\cite{pei}
This result is exact in the adiabatic limit $\Omega \rightarrow 0$
and some early works suggested that the ground state was also a Peierls
insulating state for arbitrary coupling at finite phonon 
frequency.~\cite{hir,zhe89}
More recent results suggest that quantum lattice
fluctuations can destroy the Peierls insulating state at weak
coupling or large phonon frequency.~\cite{zha,wu}
Furthermore, it is known rigorously that the ground state of a 
finite half-filled Holstein lattice is not degenerate for any 
finite electron-phonon
coupling and any finite phonon frequency.\cite{fre}
Therefore, the ground state of the Holstein model
can be a true Peierls state with CDW order and lattice
dimerization only in the thermodynamic limit.
We also note that, 
in the Holstein model of spinless electrons at half filling, 
it is well established that a metal-insulator transition occurs
at finite values of the electron-phonon coupling and phonon 
frequency.~\cite{wei,hir,ross,bur,zhe89,zhe98}
We have carried out a study of the ground state of
the one-dimensional Holstein model for electrons with spin-$\frac{1}{2}$
at half filling using the DMRG method.
Calculations of static correlation functions in long chains (up to 80 sites)
clearly show that there is a transition from a metallic ground state to a
Peierls insulating state with long range CDW order in the
thermodynamic limit.
For $\Omega = t$ this transition occurs between
$\gamma = 0.8t$ and $\gamma = 0.9t$.
In this paper we concentrate on the dynamical properties
of the six-site system and our DMRG results will be reported 
elsewhere.~\cite{jecnew}

\begin{figure}[ht]
\epsfxsize=3.375 in\centerline{\epsffile{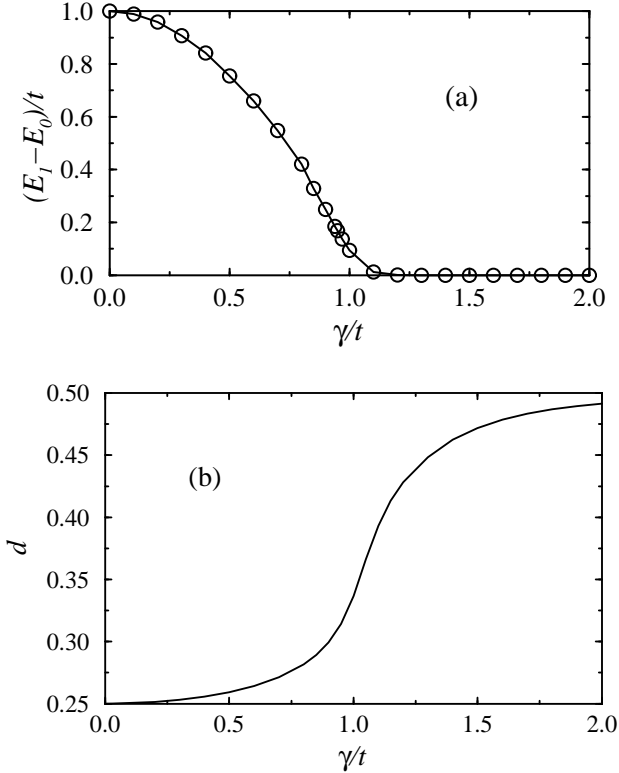}}
\caption{Half-filled system: (a) gap between the ground state
energy $E_0$ and the energy $E_1$ of the first excited state 
and (b) density of doubly occupied sites $d$
as a function of the electron-phonon coupling $\gamma$.
}
\label{fig:ED6}
\end{figure}

We had previously noticed~\cite{zha} that
static correlation functions reveal a crossover
from a uniform ground state to a Peierls CDW ground state
in a half-filled six-site Holstein lattice
around $\gamma=t$ for $\Omega=t$. 
In fact, despite the absence of a true broken symmetry ground state in a
finite Holstein lattice, signs of this crossover and the existence of an 
``insulating'' Peierls CDW phase are clearly
seen in static and dynamical properties.
This is due to a quasi-degeneracy of the ground state at strong enough
coupling.
In Fig.~\ref{fig:ED6}(a) we show the energy difference between the ground 
state $|\psi_0\rangle$ and the first excited state $|\psi_1\rangle$. 
Above $\gamma=1.1t$ the difference is very small and the two states
are almost degenerate.
The eigenstates $|\psi_0\rangle$ and $|\psi_1\rangle$
have momentum $0$ and $\pi$, respectively.
We know that they have a constant density 
$\langle \psi_{0,1} | n_i | \psi_{0,1} \rangle = 1$ 
and a uniform lattice structure 
$\langle \psi_{0,1} | b^\dag_i + b_i | \psi_{0,1} \rangle = 2\gamma/\Omega$,
because of the translation symmetry of the Holstein Hamiltonian~(\ref{eq:ham})
with periodic boundary conditions.
If they were exactly degenerate, we could build two broken symmetry 
eigenstates $|\psi_\pm\rangle = |\psi_0\rangle \pm |\psi_1\rangle$
with a charge modulation
$\langle \psi_\pm | n_i | \psi_\pm \rangle = 1 \pm (-1)^i \delta$
and a dimerized lattice 
$\langle \psi_\pm |b^\dag_i + b_i| \psi_\pm \rangle 
= 2\gamma/\Omega \; (1 \pm (-1)^i \delta)$. 
These two states would correspond to the two possible phases, 
of the degenerate Peierls CDW state obtained in mean-field 
approximation or in the adiabatic limit.~\cite{hir}
As the two-lowest eigenstates $|\psi_0\rangle$ and $|\psi_1\rangle$
are quasi-degenerate for $\gamma>1.1t$,
the six-site system properties are almost indis-

\begin{figure}[ht]
\epsfxsize=3.375 in\centerline{\epsffile{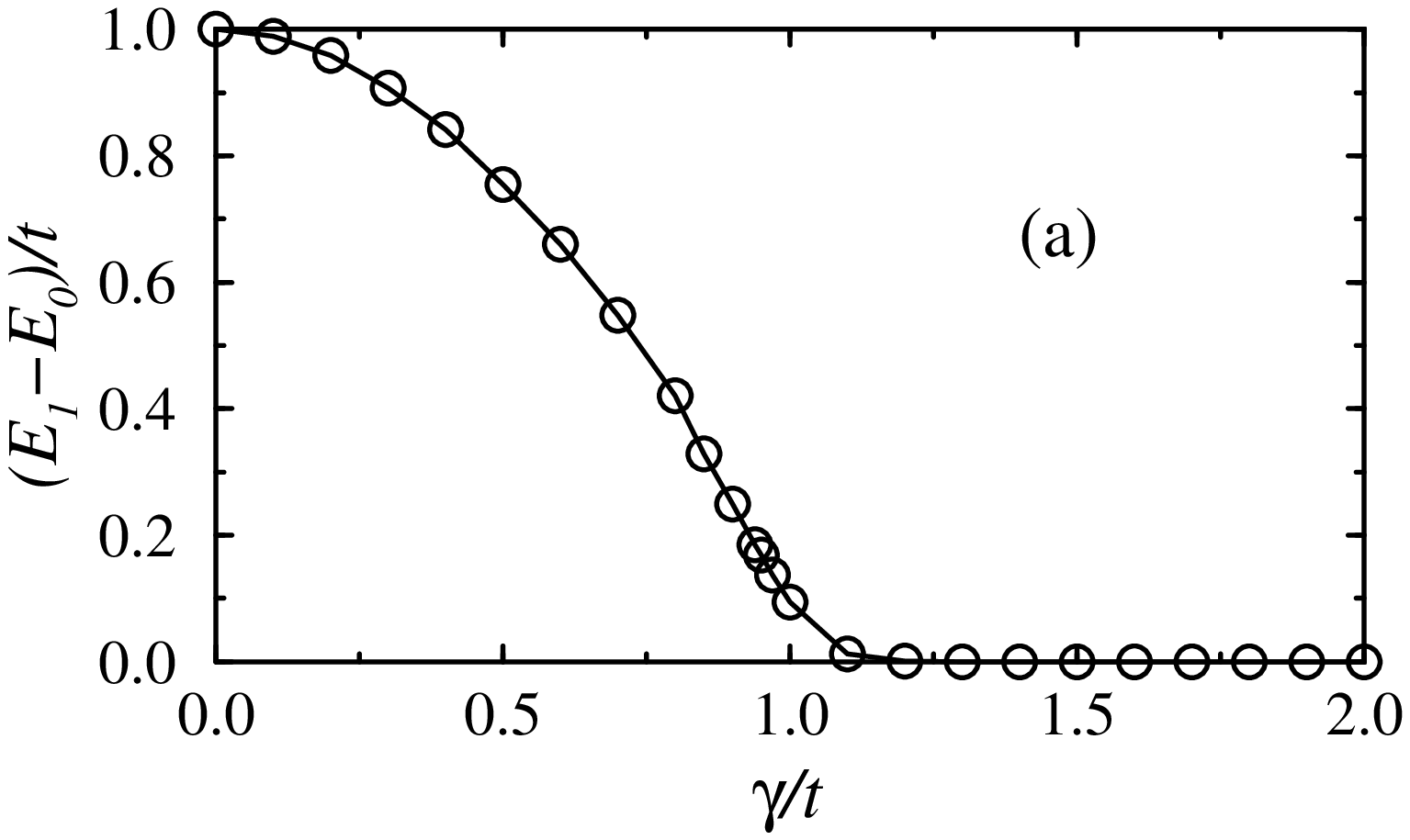}}
\end{figure}
\vspace{-8mm}

\noindent
tinguishable from those 
of a true degenerate Peierls CDW state. 

The optimal phonon wave functions for the half-filled Holstein model
have already been discussed in our previous work.~\cite{zha}
As expected, they are similar to the wave function obtained
in the two-electron system and only the relative weights of the 
optimal states are quite different due to the different number of electrons 
in the system.
The total weight of optimal phonon states associated
with zero or two electrons on a site tends to 1,
while the weight of those associated with one electron
on a site becomes very small for strong coupling.
This aspect of the crossover from independent electrons to
the Peierls CDW state is also illustrated by Fig.~\ref{fig:ED6}(b).
The density of doubly occupied electronic sites increases rather
sharply from the independent electron result $1/4$ to the
maximal value $1/2$ possible in a half-filled system.
Thus, above the crossover regime the electrons form tightly bound pairs.
These electronic pairs are heavily dressed by phonons and thus, 
can be seen as small bipolarons.~\cite{zha}
In the strong-coupling anti-adiabatic limit, it is known exactly 
that at half-filling
these small bipolarons form an ordered phase,
which is fully equivalent to the Peierls CDW state in 
this regime.~\cite{hir}

The spectral weight functions of the half-filled system are harder
to interpret than those of the single- and two

\begin{figure}[ht]
\epsfxsize=3.375 in\centerline{\epsffile{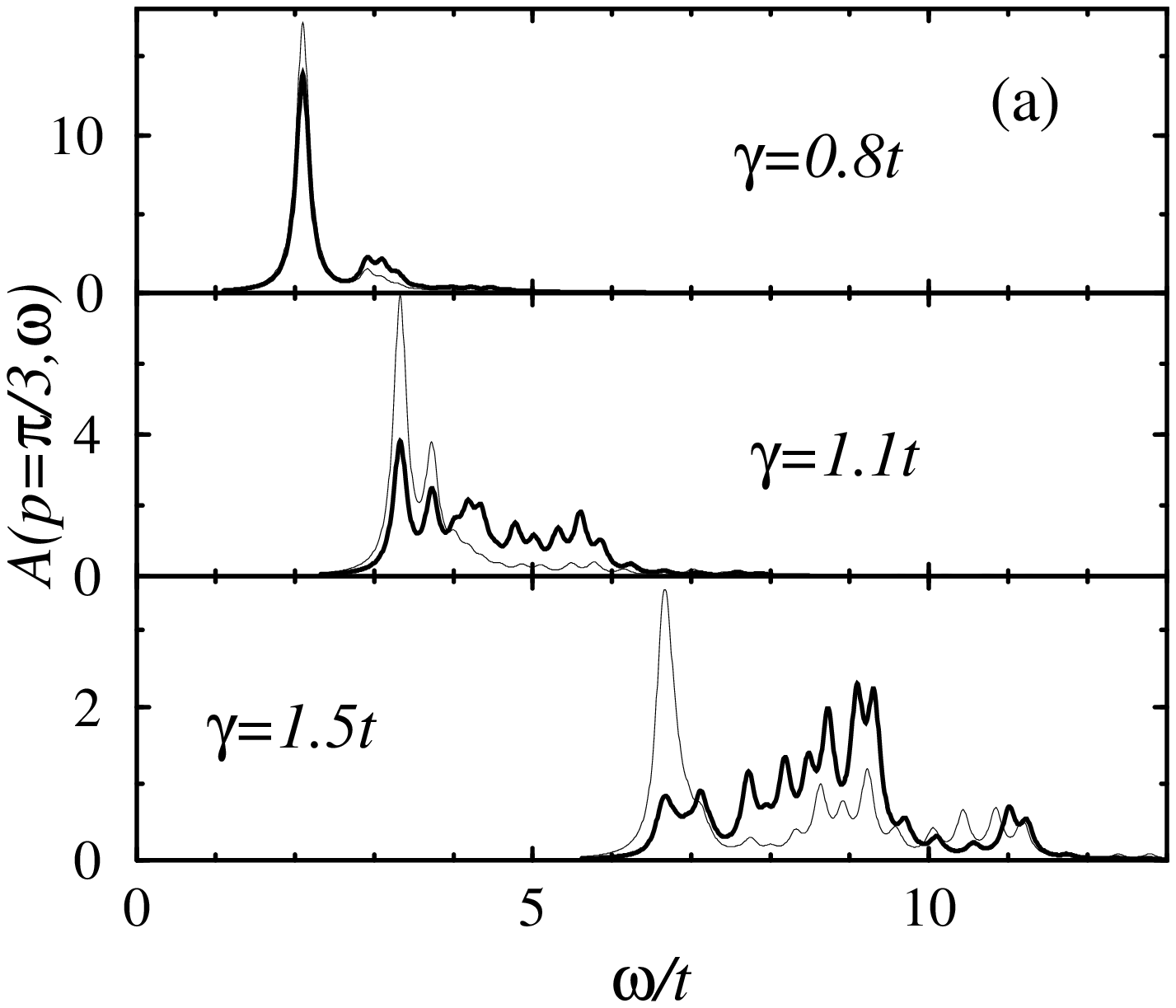}}
\end{figure}
\vspace{-7mm}
\begin{figure}[ht]
\epsfxsize=3.375 in\centerline{\epsffile{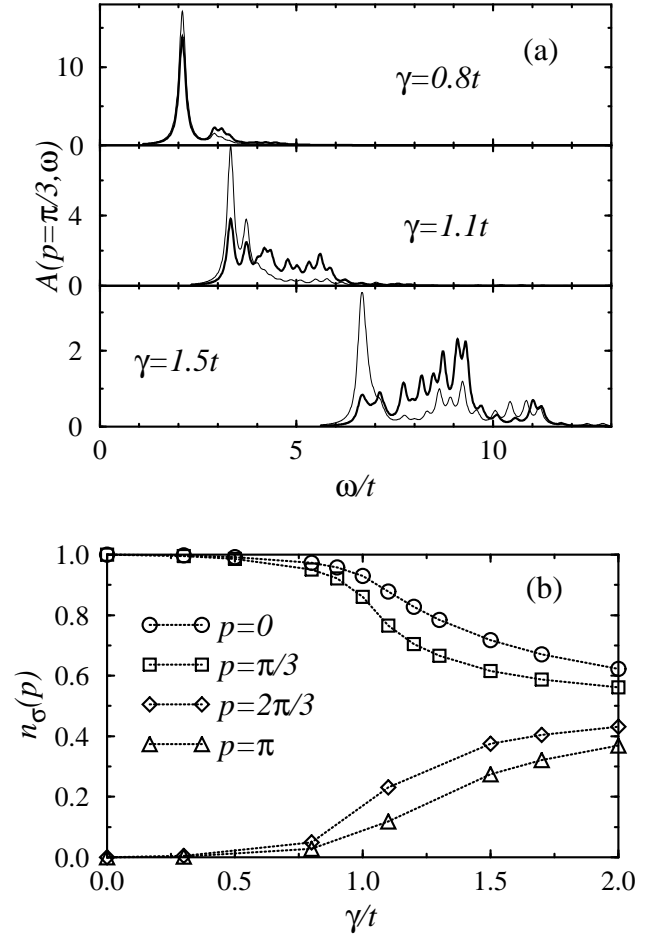}}
\caption{Half-filled system: (a)
the spectral functions $A(p=\pi/3)$ (thick line) and
$\tilde{A}(p=\pi/3)$ (thin line) 
for different electron-phonon couplings $\gamma$.
(b) The momentum density distribution 
$n_\sigma(p)$ for all values of the momentum $p$ 
(results for $p$ and $-p$ are identical)
as a function of the electron-phonon coupling.
}
\label{fig:A6}
\end{figure}

\noindent
-electron system. 
Nevertheless, some of their properties can be understood.
In the non-interacting limit, the spectral functions
$A(p,\omega)$ and $\tilde{A}(p,\omega)$ show a single peak
with total weight 1 for momentum $|p| \leq \pi/3$ 
corresponding to occupied single-electron states
and are uniformly
zero for momentum $|p| > \pi/3$ corresponding to
unoccupied single-electron states.
As the electron-phonon coupling increases, both
$A(p,\omega)$ and $\tilde{A}(p,\omega)$
become fairly complex for all values of $p$.
As an example, Figure~\ref{fig:A6}(a) shows
both functions for $p=\pi/3$.
In the strong-coupling regime ($\gamma = 1.5t$),
one can see that the highest peak of the bare spectral function
$A(p,\omega)$ is located at an energy
$\omega \approx 4\gamma^2/\Omega$ corresponding to the energy
required to remove one electron of the half-filled system
without perturbing the dimerized lattice structure of the Peierls CDW 
ground state.
On the other hand, the spectral weight for dressed
electrons $\tilde{A}(p,\omega)$ shows a relatively sharp dominant peak,
but with only a fraction of the total weight,
for all couplings.
The position of this peak 

\begin{figure}[ht]
\epsfxsize=3.375 in\centerline{\epsffile{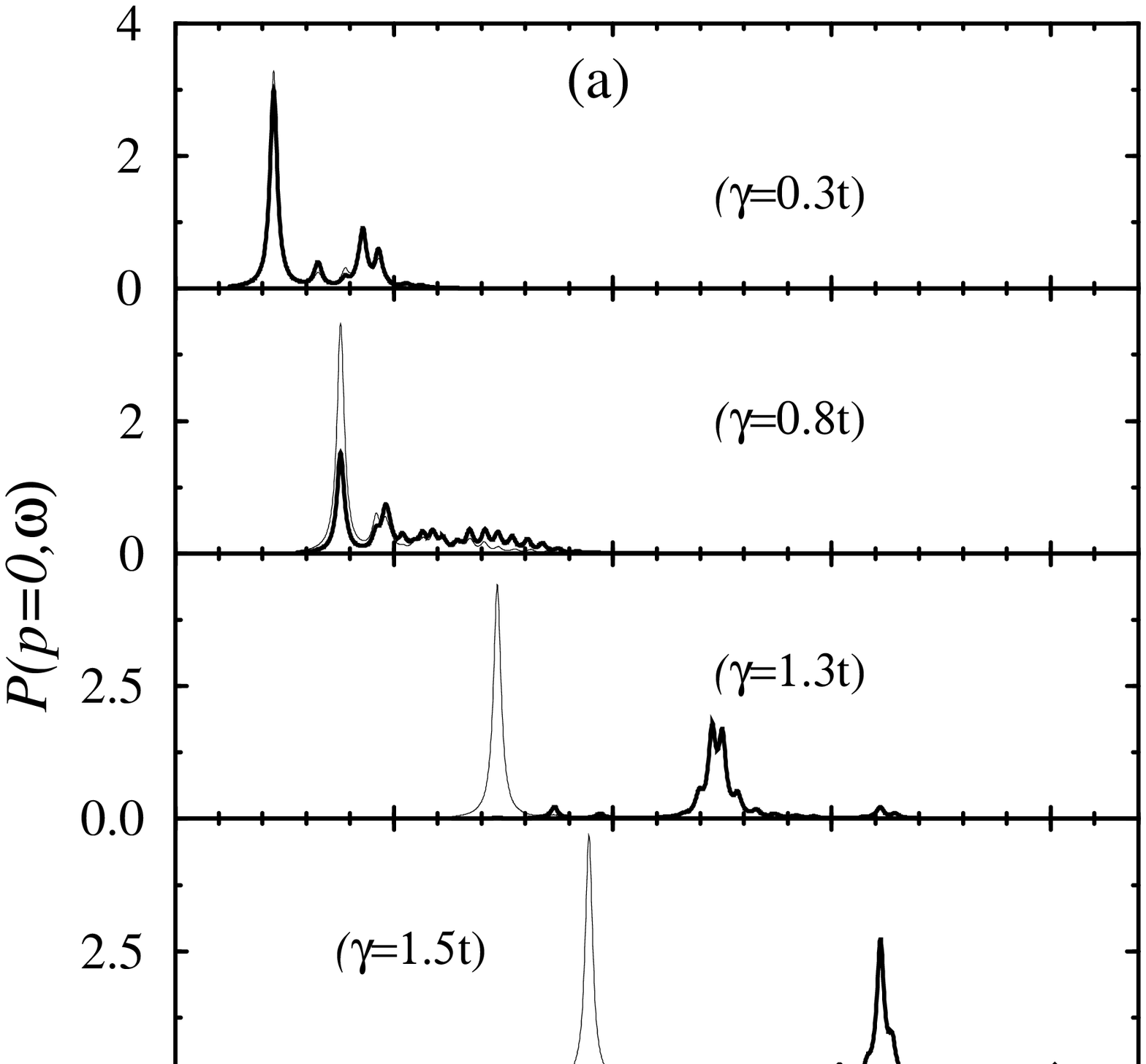}}
\end{figure}
\vspace{9mm}
\begin{figure}[ht]
\epsfxsize=3.375 in\centerline{\epsffile{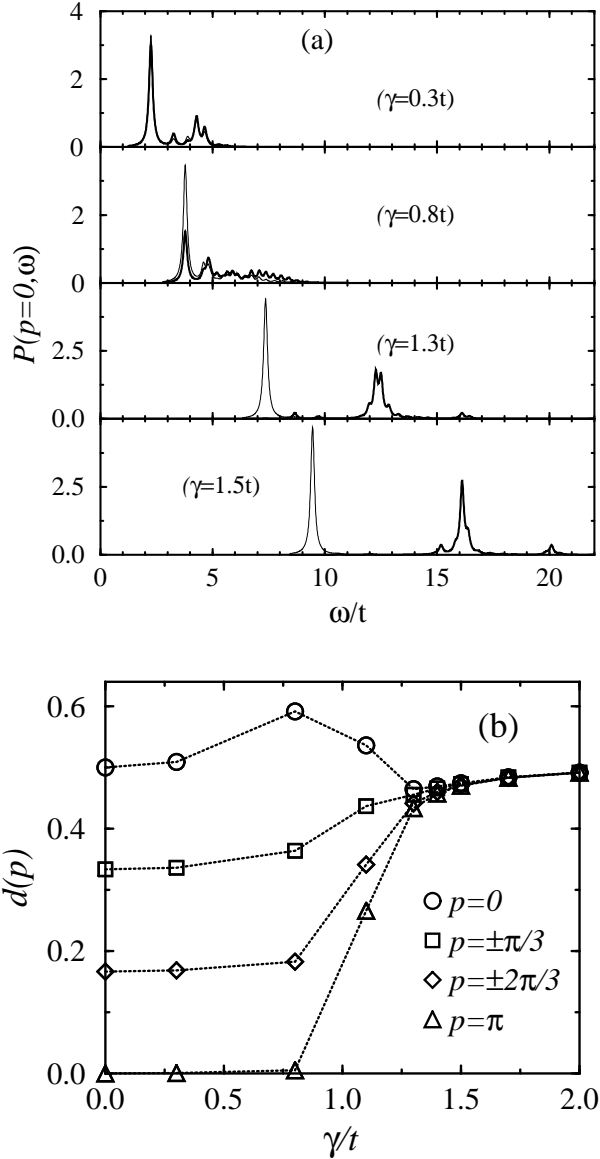}}
\caption{Half-filled system: (a) the pair spectral functions 
$P(p=0,\omega)$ (thick line) and $\tilde{P}(p=0,\omega)$
(thin line) 
for different electron-phonon couplings $\gamma$.
(b) The total spectral weight $\tilde{d}(p)$ for
all values of the momentum $p$
as a function of the coupling $\gamma$.
}
\label{fig:P6}
\end{figure}

\noindent
corresponds to the difference between
the ground state energy of the half-filled system and that
of a system with one electron removed from a half-filled band.
For $\gamma=1.5t$ we find that this peak is around 
$\omega = 3\gamma^2/\Omega$
in agreement with strong-coupling theory predictions.
The total spectral weight $n_\sigma(p)$ is shown in Fig.~\ref{fig:A6}(b)
as a function of $\gamma$.
Results for $\tilde{n}_\sigma(p)$ are similar.
One can see that $n_\sigma(p)$ evolves smoothly from the
momentum density distribution of free electrons,
$n_\sigma(p) = 1$ for $|p| \leq \pi/3$ and $n_\sigma(p) = 0$ for
$|p| > \pi/3$, to that of completely
localized electrons, $n_\sigma(p) = 0.5$ for all $p,\sigma$.

At weak coupling, the properties of the pair spectral 
functions $P(p,\omega)$ and $\tilde{P}(p,\omega)$
depend strongly on the momentum $p$ but can be easily understood
from weak-coupling perturbation theory.
In the Peierls CDW regime both functions acquire a simple structure,
which is largely similar for all values of the momentum $p$.
For instance, Figure~\ref{fig:P6}(a) shows both functions
for $p=0$ from  weak to strong coupling.
For very weak coupling the bare pair spectral function
$P(p=0,\omega)$ has only two peaks at $\omega=2t$ and $4t$,
which correspond to zero momentum pairs made of two electrons
with momentum $p=\pm\pi/3$ and $p=0$, respectively.
As the coupling increases toward the crossover regime, the
spectral weight is shifted to an increasing number of
satellite peaks. 
In the Peierls CDW regime we observe a relatively broad cluster of peaks.
This cluster seems to be centered around an energy
which is slightly lower than the energy $8\gamma^2/\Omega$
required to remove two
bare electrons from the half-filled ground state
without disturbing the dimerized lattice structure.
For all couplings $\tilde{P}(p=0,\omega)$ has a well-defined
dominant peak, although incoherent contributions are significant
in the quasi-free-electron regime.
The position of this peak gives the energy difference between the
half-filled band ground state and the ground state of
a system with two electrons removed from a half-filled band.
For $\gamma=1.5t$ one can see that this energy difference
is about $4\gamma^2/\Omega$ in agreement with strong-coupling theory.
As noted above similar results are obtained for all other momentum
in the Peierls CDW phase.

The well-defined peaks observed in $\tilde{P}(p,\omega)$
confirm that the ground state is made of small bipolarons
in this regime.
In such a case, the total weight $\tilde{d}(p)$ can be seen
as the bipolaron momentum distribution.
The distribution $\tilde{d}(p)\approx 0.5$ for all momentum $p$
that we observe at strong-coupling [see Fig.~\ref{fig:P6}(b)]
indicate that bipolarons are completely localized in the Peierls CDW regime.
The study of static correlation functions shows that these localized 
small bipolarons form an ordered phase even for finite electron-phonon 
coupling and finite phonon frequency.~\cite{zha}
Therefore, we think that the ground state of the half-filled six-site 
Holstein lattice
can be seen either as a Peierls CDW state with lattice dimerization or
as an ordered phase of localized small bipolarons.
The first point of view corresponds to the adiabatic limit
($\Omega/t\rightarrow 0$) result and the second is more appropriate
in the strong-coupling or anti-adiabatic limit ($t/\Omega\rightarrow 0$).
For the intermediate case $\Omega=t$ discussed here, both
pictures appear completely equivalent.

\begin{figure}[ht]
\epsfxsize=3.375 in\centerline{\epsffile{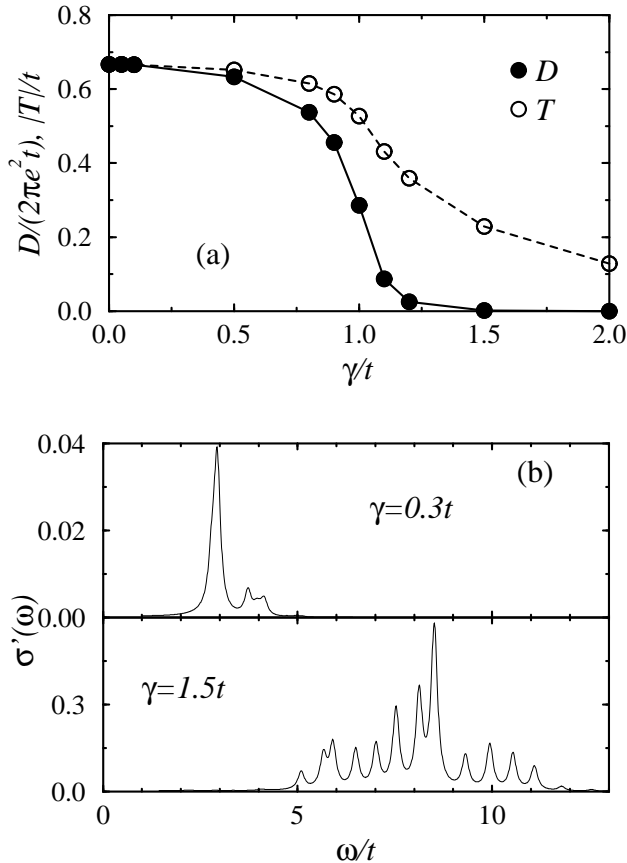}}
\caption{Half-filled system: (a) the Drude weight $D$ and 
the kinetic energy per site $T$ as a function of the electron-phonon 
coupling $\gamma$.
(b) The incoherent part of the optical conductivity $\sigma'(\omega)$
in the quasi-free-electron regime ($\gamma=0.3t$) and in
the Peierls CDW regime ($\gamma=1.5t$).
}
\label{fig:S6}
\end{figure}

Figure~\ref{fig:S6}(a) shows the evolution of the Drude weight and of
the kinetic energy per site as the electron-phonon coupling increases.
Again the units are chosen so that both
quantities are equal in the absence of incoherent processes in the
optical conductivity.
The kinetic energy decreases rather smoothly as in the 
single- and two-electron cases.
On the other hand, we can clearly see a sharp decrease of the 
Drude weight around $\gamma=t$.
Above $\gamma=1.5t$, $D$ vanishes within numerical errors.
(Of course, $D$ is never really zero on a finite cluster because
of quantum tunneling.)
This negligible value of $D$ is consistent 

\begin{figure}[ht]
\epsfxsize=3.375 in\centerline{\epsffile{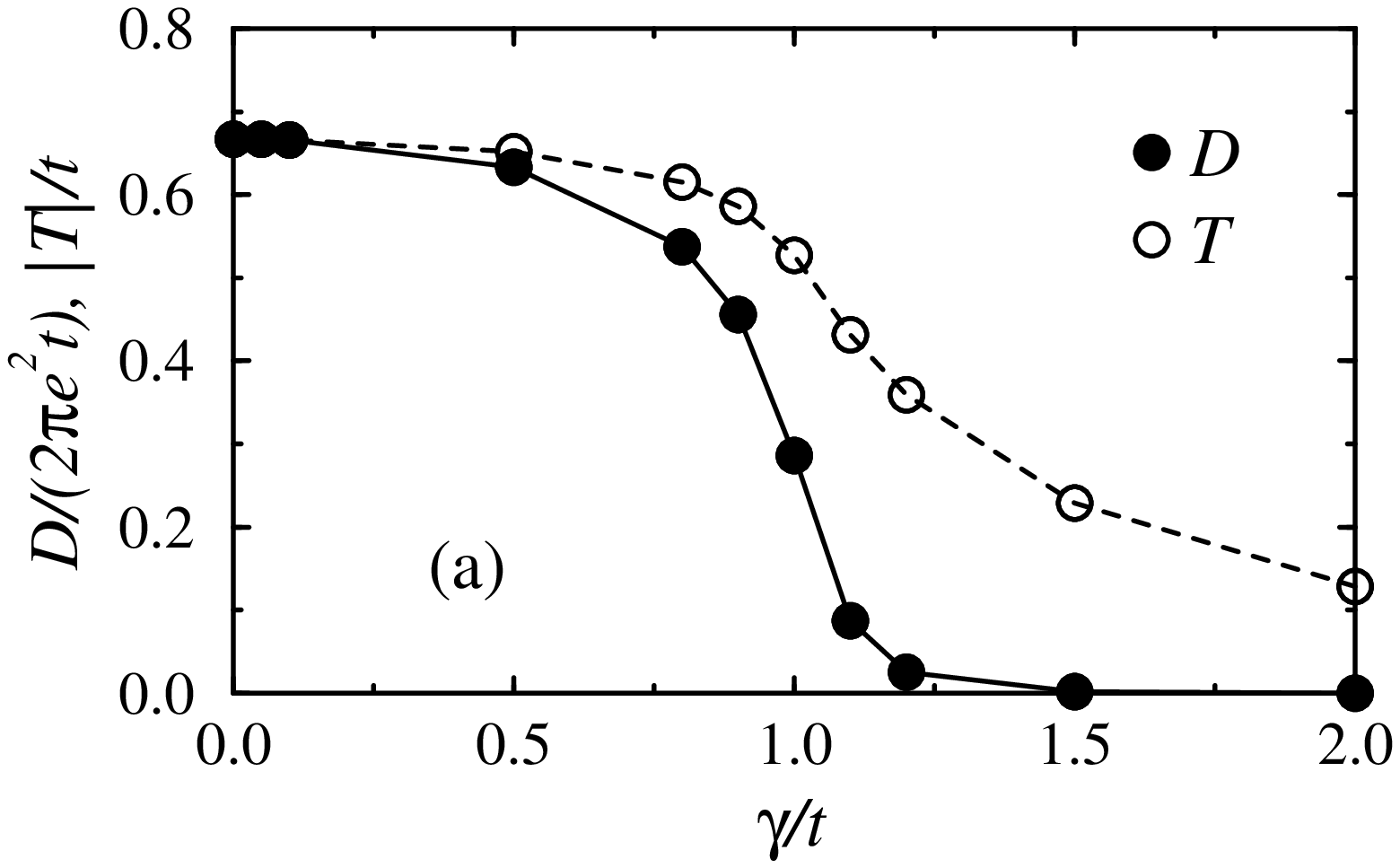}}
\end{figure}
\vspace{-5mm}

\noindent
with the ground state
being made up of localized small bipolarons.
Although a  six-site cluster can not be metallic or insulating,
the behavior of $D$ shown here illustrates perfectly the 
metal-insulator transition observed in the
thermodynamic limit.~\cite{jecnew}
The incoherent part of the optical conductivity is shown 
in Fig.~\ref{fig:S6}(b).
For weak coupling ($\gamma=0.3t$), 
$\sigma'(\omega)$ has very little
weight and contains a single dominant peak that can be explained by the
discrete electronic energy levels of the non-interacting system.
In the Peierls CDW regime ($\gamma=1.5t$), 
the structure of $\sigma'(\omega)$
is more complex but clearly shows a significant peak around an energy
$\omega \approx 4\gamma^2/\Omega$ corresponding to the 
Peierls gap in the strong-coupling limit.
In the adiabatic limit the conductivity is zero below the Peierls gap.
The significant tail observed below the peak in Fig.~\ref{fig:S6}(b)
is due to phonons.
It is an excellent illustration
of the subgap optical absorption predicted in Peierls
systems because of quantum lattice fluctuations.~\cite{ken}

Our results clearly show that there are two distinct regimes
in the half-filled six-site Holstein lattice.
The crossover from one regime to the other occurs rather sharply at a 
critical electron-phonon coupling, which is $\gamma\approx t$ for
$\Omega=t$.
Below this critical coupling, the static and dynamical properties
of the system are those of quasi-free electrons on a finite cluster.
Above the critical coupling, the ground state is made of ordered
localized small bipolarons, which can also be seen as the Peierls
state with CDW order and lattice dimerization predicted
by mean-filed theory.
These two regimes are the finite system precursors of the
metallic and insulating Peierls CDW phase of the infinite system,
and the crossover between both regimes
signals the quantum phase transition 
observed at finite electron-phonon coupling 
in the thermodynamic limit of the half-filled Holstein model.~\cite{jecnew}

\section{Conclusion}

We have studied the dynamical properties of the six-site 
Holstein model for three different electron concentrations
using an exact diagonalization technique.
A density matrix approach allows us to generate an optimal
phonon basis and to truncate the phonon 
Hilbert space without significant loss of accuracy.
With this very efficient method we are able to observe
the evolution of the system properties as one goes from
the weak electron-phonon coupling regime to the strong-coupling 
regime.
For all three electron concentrations studied a smooth crossover
is observed from quasi-free electrons at weak coupling
to a strongly correlated state in the strong-coupling regime.
This strongly correlated state is
a heavy itinerant small polaron in the single-electron case,
a heavy itinerant small bipolaron in the two-electron case,
and a set of ordered localized small bipolarons similar
to a Peierls CDW state in the half-filled band case.

The study of the optimal phonon states reveal that they often are
the eigenstates of a quantum oscillator with a shifted equilibrium
position. These states can be obtained by applying the 
Lang-Firsov transformation to the bare phonon states.
The amplitude of the shift is generally smaller than the
exact results for $t=0$ but tends to this value for 
strong coupling.
However, in some cases we have observed some 
retardation effects due to the electron motion and
the slow response of the phonon modes.
These effects are small but they are essential for an accurate 
description of the electron motion along the lattice, 
either as independent particles or as part of a composite quasi-particle
(polaron or bipolaron). 
Although we have presented only results for $\Omega=t$,
we have checked that these retardation effects become
more important for smaller values of $\Omega/t$ (adiabatic limit)
but vanish for larger values of $\Omega/t$ (anti-adiabatic limit).

We have obtained a wealth of information from the single-particle and 
pair spectral weight functions.
By dressing the electron operators with the optimal phonon states,
we are often able to simplify the structure of these spectral functions
and to obtain well-defined quasi-particle peaks.
For instance, we can identify the formation of a small polaron and
of a small bipolaron by the appearance of a single dominant peak in 
the single electron spectral function $\tilde{A}(\omega)$ 
and in the pair spectral function $\tilde{P}(\omega)$, 
respectively.  
We have also studied the optical conductivity for these systems.  
In all cases the Drude weight decreases substantially
as the electron-phonon coupling increases from the non-interacting limit 
to the strong-coupling limit.
At half filling, the Drude weight decreases abruptly around $\gamma=t$ and
is negligibly small for larger $\gamma$. 
This suppression of coherent transport is linked to the appearance
of a quasi-degenerate Peierls CDW ground state.
These results support measurements of static correlation functions in
larger systems which shows
the existence of a metal-insulator transition in the thermodynamic limit
of the half-filled one-dimensional Holstein model. 

One obvious limitation of this study is the complete
neglect of electron-electron interaction when there is
more than one electron on the lattice.
This interaction is likely to strongly affect the properties
of the system, especially in the regime where bipolarons are formed
in the absence of electronic repulsion.
The approach used in this work can be applied without difficulty
to models with an electron-electron interaction and we have started
to investigate the Holstein-Hubbard model,
which includes an on-site electron-electron repulsion.

\section{Acknowledgments}
We would like to thank A. Weisse and H. Fehske for putting some 
of their results at our disposal before publication.
E.J. thanks the Institute for Theoretical Physics of the University of 
Fribourg, Switzerland, for its kind hospitality during the preparation 
of this paper.
S.R.W. acknowledges support from the NSF under
Grant No. DMR-98-70930, and from the University of California
through the Campus Laboratory Collaborations Program.

\end{document}